\documentclass[prd,groupaddress,nofootinbib]{revtex4}
\usepackage{graphicx}
\usepackage{amssymb,amsmath,amsfonts}
\def\0{\nonumber}

\def\dag{\dagger}
\def\del{\partial}

\def\g{\gamma}        
\def\d{\delta}        
\def\e{\epsilon}

\def\m{\mu}	   \def\M{M}        
\def\n{\nu}                
\def\x{\xi}

\def\s{\sigma}

\def\y{\psi}

\def\br{\langle}
\def\ke{\rangle}
\def\ve{\vert}
\def\inf{\infty}
\def\Winf{$W_{\infty}\  $}
\def\winf{$w_{\infty}\  $}
\def\to{\rightarrow}

\def\zbar{\bar{z}}
\def\zintm{d^2 z e^{-|z|^2}}

\def\to{\rightarrow}
\def\tr{{\rm tr}}

\def\ba{{\cal A}}

\newcommand{\ET}{{T}}

\newcommand{\be}{\begin{equation}}
\newcommand{\ee}{\end{equation}}
\newcommand{\bea}{\begin{eqnarray}}
\newcommand{\eea}{\end{eqnarray}}
\newcommand{\nn}{\nonumber\\}

\def\lsim{\hbox{ \raise.35ex\rlap{$<$}\lower.6ex\hbox{$\sim$}\ }}
\def\gsim{\hbox{ \raise.35ex\rlap{$>$}\lower.6ex\hbox{$\sim$}\ }} 

\begin{document} 

\title{$W_\infty$ Algebras, Hawking Radiation and Information Retention by Stringy Black Holes\\
~\\}

\author{John Ellis} 

\author{Nick E. Mavromatos}

\affiliation{Theoretical Particle Physics and Cosmology Group, Department of Physics, \\
King's College London, Strand, London WC2R 2LS, U.K; \\
Theoretical Physics Department, CERN, CH 1211 Geneva 23, Switzerland}

\author{D. V. Nanopoulos}

\affiliation{George P. and Cynthia W. Mitchell Institute for Fundamental Physics and Astronomy,\\
Texas A\&M University, College Station, TX 77843, USA; \\
Astroparticle Physics Group, Houston Advanced Research Center (HARC),
Mitchell Campus, Woodlands, TX 77381, USA;\\
Academy of Athens, Division of Natural Sciences, Athens 10679, Greece}

\begin{abstract}
\vspace{0.1in}

We have argued previously, based on the analysis of two-dimensional stringy black holes,  
that information in stringy versions of four-dimensional Schwarzschild black holes 
(whose singular regions are represented by appropriate Wess-Zumino-Witten models)
is retained by quantum $W$-symmetries when the horizon area is not preserved due to Hawking radiation.
It is key that the exactly-marginal conformal world-sheet operator representing a massless stringy particle interacting with the 
black hole requires a contribution from $W_\infty$ generators
in its vertex function. The latter correspond to delocalised, non-propagating, string excitations that
guarantee the transfer of
information between the string black hole and external particles. When infalling matter crosses the horizon,
these topological states are excited via a process:
(Stringy black hole) + infalling matter $\rightarrow $  (Stringy black hole)$^\star$,
where the black hole is viewed as a stringy state with a specific configuration of 
$W_\infty$ charges that are conserved. Hawking radiation is then the reverse process,
with conservation of the $W_\infty$ charges retaining information.
The Hawking radiation spectrum near the horizon of 
a Schwarzschild or Kerr black hole is specified by matrix elements of higher-order currents
that form a phase-space $W_{1+\infty}$ algebra. We show that an appropriate gauging 
of this algebra preserves the horizon two-dimensional area classically,
as expected because the latter is a  conserved Noether charge.\\
~\\

\begin{centering}
~~~~~~~~~~KCL-PH-TH/2016-23, LCTS/2016-10, CERN-TH-2016-105, ACT-04-16, MI-TH-1615
\end{centering}
\end{abstract}

\maketitle

\section{Introduction and Summary}

The black-hole information problem was posed by the discoveries by
Bekenstein~\cite{bek} and Hawking~\cite{hawk} that four-dimensional black holes have
thermodynamical properties such as temperature and non-zero entropy,
and so must be described by mixed quantum-mechanical
states. These discoveries led Hawking~\cite{hawk2}, in particular,
to suggest that information would be lost across
the black-hole horizon, giving rise to a transition from a pure to a mixed state.

String theory has provided an explicit theoretical laboratory for probing the black-hole information problem, 
notably using the two-dimensional black-hole solution found by Witten~\cite{witt}
that has an SU(1,1)/U(1) coset structure~\cite{horava,Chlyk},
coupled with dualities~\cite{KN}, and subsequently using four-dimensional stringy black holes constructed using
D-branes~\cite{dbranebh,zanon}. We argued~\cite{emn1} that two-dimensional
black holes carry an infinite set of quantum numbers associated with a $W_\infty$ symmetry,
and that these $W$ charges preserve the lost information {\it in principle}, 
though this information could not {\it in practice} be extracted.
These observations apply also
to spherically-symmetric four-dimensional stringy black holes~\cite{emn2}, which have horizons whose
geometry is also encoded in an SU(1,1)/U(1) coset structure that possesses
a similar $W_\infty$ symmetry and an associated infinite set
of $W$ `hair'~\cite{wmeasure}.
 
D-brane constructions provide examples of four-dimensional
black-holes whose microstates could be counted explicitly~\cite{dbranebh}, with results
consistent with the Bekenstein-Hawking entropy,
suggesting again that the `lost' information was retained {\it in principle}.
However, there were still questions how this information was transferred
from external particles to these microstates, how it was stored,
and whether the information retained by the microstates could {\it in practice}
be extracted from observations of radiated particles.

An interesting, complementary approach to the black-hole information problem has been taken recently
by Strominger and collaborators~\cite{strominger,strominger2}. They have argued that
spherically-symmetric four-dimensional black holes carry a previously undiscussed infinite set
of soft gravitational `hair' associated with BBMS supertranslations and superrotations~\cite{bms,bms2,HPS} on the retarded
null infinity $\cal{I}^+$. These correspond to vacua that differ by the addition of
soft gravitons that could {\it in principle} be measured via the gravitational memory effect.
Hawking, Perry and Strominger (HPS)~\cite{HPS} also discusses in detail
an analogous infinite set of inequivalent electromagnetic gauge
configurations corresponding to soft electromagnetic `hair' that differ by the
addition of soft photons and could also be distinguished {\it in principle} by measurements~\footnote{See, however,
\cite{Herdegen:2016bio} for a critical discussion and references to earlier work.}.
Assuming additionally that
soft hair localized spatially to much less than the Planck length would not be excited in a physical
process, HPS found~\cite{HPS}
an effective number of soft degrees of freedom proportional to the horizon area,
like the Hawking-Bekenstein entropy. 
At the present stage of this proposal, it is unclear whether supertranslations and the corresponding
superrotations and electromagnetic gauge configurations can
encode the information carried by the incoming particles~\cite{dvali}, 
and HPS did not claim to have resolved the information paradox.
Moreover, the connection between this approach and one based on stringy black holes is not apparent, 
and it is the latter approach that we adopt in this paper.

In contrast to the recent work of `t Hooft~\cite{thooft}, we base our work on string theory at a fundamental level.
Our analysis below of supertranslations on the 
horizon, viewed as the recoil of a D-brane induced by infalling matter, is more similar in spirit to
the recent work of Polchinski~\cite{polch}, in which a shock-wave approximation was used to 
calculate the shift of a generator of the horizon caused by an incoming wave packet.
We recall also the work of~\cite{englert,brus} where the fluctuation of the black hole horizon induced by infalling matter
is argued to play an important  r\^ole in retaining information. In this connection, 
we recall that in string theory the interaction of a massless particle with a 
black hole is represented by a conformal operator on the world-sheet of the string,  which is exactly marginal 
{\it if and only if} contributions from $W_\infty$ generators are included in its vertex function~\cite{emnQM,Chlyk}.
The corresponding renormalization-group (RG) $\beta$ function would be non-zero without these contributions,
leading to a monotonic increase in entropy. 

As reviewed recently in~\cite{emn2015}, we consider $W_{1+\infty}$ symmetry~\cite{iso,wcontr}
to be essential for `balancing the black-hole information books'. This symmetry is manifest in the 
effective two-target-space dimensional string theories that describe the excitations in the near-horizon geometry 
of a spherically-symmetric stringy black hole. It is larger than the symmetry group of
supertranslations, and we consider that the latter are insufficient to retain fully the black-hole information.
The purpose of this article is to study further the r\^ole of these $W_\infty$ symmetries in retaining information,
and to establish a connection with the work of~\cite{bonora,higherspin}, 
in which the spectrum of Hawking radiation is related to matrix elements of such an
infinite-dimensional $W_\infty$ algebra. In this connection,
we recall~\cite{emn1,emnQM} that $w_\infty$ (the classical limit of the quantum $W_\infty $ symmetry)
is the algebra of transformations that preserve the two-dimensional phase-space 
volume of massless (`tachyonic') stringy matter propagating in the background of a stringy black hole.  

The quantum version of this $w$ algebra is a symmetry of the quantum scattering matrix of the 
corresponding two-dimensional string theory~\cite{klebanov,Chlyk}, in the sense that the 
operator product expansion between two appropriate vertex operators reproduces
the corresponding $W$ algebra. In the flat space-time case (where the 
string theory is just a two-dimensional Liouville theory), the operators corresponding to the 
discrete higher-spin operators of the $W$ algebra are discretized `tachyon' operators. 
However, as already mentioned, in the presence of a black hole, at the quantum level
the corresponding $W_\infty$ symmetries necessarily mix massless and 
massive stringy states that are topological and delocalised~\cite{Chlyk}. The admixture of $W_\infty$ generators
in the exactly-marginal vertex operator of a massless string excitation makes manifest the transfer of
information between a stringy black hole and external particles.

In addition to its spectrum, another important feature of the Hawking radiation is its sparsity at 
asymptotic infinity. This feature can be explained by viewing the black holes as `particles' and the 
Hawking radiation process as successive two-body decays~\cite{visser}. As we shall see, 
such a picture emerges naturally in string theory, but with essential differences,
since the black holes are represented as string states that are completely integrable, 
due to their infinity of conserved $W_\infty$-charges.

In this article we synthesise these ideas from a current perspective.
In Section~\ref{sec:w} we review the properties and formalism underlying two-dimensional stringy black holes and their 
embedding in four-dimensional space times, using them to represent spherically-symmetric black-hole configurations in 
four space-time dimensions. The underlying two-dimensional coset structure of the singularity is essential for the 
\emph{complete integrability} of these systems and the retention of information via the corresponding 
$w_\infty$ space-time symmetry current algebras that characterise them. A review of these symmetries 
and the construction of the corresponding currents in terms of discrete higher-spin string states is also given in this Section. 
We also review in this context our approach to quantifying information loss
by representing entropy increase in an evaporating black hole in string theory as 
world-sheet renormalization group flow between fixed points in the stringy world-sheet field theory space. 
We also mention briefly the case of a string-inspired infinitely-coloured SU($\infty$)  black hole~\cite{winstanley}, 
where the conserved entropy is linked classically to a $w_\infty$ symmetry living on the horizon and preserving
its area~\cite{fit,ilio,emn2015}. This property is consistent with viewing the classical black-hole area 
as a conserved Noether charge~\cite{wald}. In Section~\ref{sec:hr} we discuss Hawking radiation in 
generic four-dimensional spherically-symmetric black holes and its link~\cite{bonora} with
another form of $W_{1+\infty}$ phase-space symmetry algebra associated with higher-spin currents  
that correspond to fluxes of Hawking radiation in the effective two-dimensional field theory 
representation of the Hawking radiation thermal spectrum in the near-horizon geometry. 
We compare the situation with our stringy case, which involves non-thermal higher-spin 
discrete delocalised states, also associated with phase-space $W_{1+\infty}$ algebras.  
In Section~IV we discuss the gauging of this algebra and its connection with the preservation of
the horizon area at the classical level, interpreted as the conservation of a Noether charge, with some technical details provided in the Appendix.
Our conclusions are presented in Section~\ref{sec:concl} where, in view of the r\^ole of 
$W$ symmetries in preserving information during matter infall or Hawking  radiation, the evaporation
of the stringy black hole is viewed as successive two-body decays. 

\section{Stringy Black Holes, $W_{1+\infty}$ Symmetry Algebras and Information Retention \label{sec:w}} 

\subsection{Two-Dimensional Stringy Black Holes as Prototypes} 

The stringy black hole in two dimensions~\cite{witt} can be formulated as
a world-sheet Wess-Zumino-Witten (WZW) $\sigma$-model on the coset space SL($2,R$)$_k$/U(1),
where $k=9/4$ is the Kac-Moody algebra level.
The two-dimensional target-space metric induced by the conformal invariance condition for this world-sheet $\sigma$-model 
corresponds to a Euclidean black-hole background
\be\label{2bh}
ds^2 = dr^2 + {\rm tanh}^2 \, r \, d \tilde \theta^2 \, ,
\ee
where $r$ is a radial coordinate and
$\tilde \theta $ is a compact `angular' coordinate that plays the r\^ole of an external temperature variable.
The Euclidean space-time (\ref{2bh}) looks like a semi-infinite cigar. 

Far away from the singularity, this space-time is asymptotically flat, and the corresponding string theory
is a $c=1$ two-target-dimensional (2D) $(X,\rho)$ Liouville theory with a space-like Liouville mode $\rho$ 
and a central charge deficit $Q=2\sqrt{2}$. In this asymptotic theory, the tachyon field operator
is the standard `massless matter' operator of the 2D Liouville theory~\cite{klebanov}: 
\be\label{tachflat}T^\pm (p) = e^{ip X + (\pm |p| -\sqrt{2}\, \rho},
\ee 
which describes a positive-norm physical state. The spectrum also contains
discrete higher-spin $J =0, \frac{1}{2}, 1, 2, \dots $ states with third component of the internal 
angular momentum $M= \{ -J, -J +1, \dots , J-1, J \}$, which also have positive norm,
with vertex operators having the asymptotic forms
\be\label{dstates}
\psi^{(\pm)}_{J, M} \sim (H_-)^{J-M} \, \psi_{J,J}^{(\pm)} \sim (H_+)^{J+M} \, \psi_{J,-J}^{(\pm)}~, \quad H_{\pm} = \int \frac{dz}{2\pi \, i} T^+(\pm \sqrt{2})~, 
\ee
where $\sim$ denotes a normalization factor, $H_\pm$ represents the zero modes of the ladder operator 
of the SU(2) Kac-Moody currents at the self-dual radius of the $c=1$ conformal field theory, and 
$\psi_{J,\pm J}^{(\pm)} = T^{(\pm)}(\pm \sqrt{2}J)$. 

The operator products (OPs) of these asymptotic discrete states 
form a classical $w_\infty$ algebra~\cite{klebanov}:
\be\label{walgebra}
\int \frac{dz}{2\pi \, i} \, \psi^+_{J_1, \, M_1}(z) \, \psi^+_{J_2, \, M_2}(z) (0) = (J_2\, M_1 - J_1, \, M_2) \, \psi^+_{J_1 + J_2 -1, \, M_1 + M_2 }(0)  + \dots \, ,
\ee
where the $\dots$ are explained below. This symmetry algebra leads to an infinity of conserved currents and charges,
and can be used to construct the (non-local) effective action in the $c=1$ 2D target-space 
Liouville theory~\cite{klebanov,pol} and its matrix-model extension~\cite{wadia}. The classical $w_\infty$ symmetry
is elevated at the quantum level to a $W_{1+\infty}$ algebra with a central extension, 
where the extra subscript ``1'' is due to the inclusion of a U(1) spin-one current. 

The $c=1$ theory has also \emph{zero-norm (ghost-like) discrete gauge states} (DGS), which, 
as explained in \cite{lee}, also satisfy the physical-state Virasoro-operator conditions, like the positive-norm discrete states,
and represent the same $W_{1+\infty}$ algebra as the positive-norm states. 
These gauge states have discrete momentum values corresponding exactly to the physical 
positive-norm discrete states. Detailed formulae for these discrete states are given in \cite{lee} and will not be repeated here. 
For information, we give the expression for one class of these states, in the $\psi^+$ sector:
\be\label{dgs}
G_{J,M}^+ = (J + M +1)^{-1} \int \frac{d z}{2\pi \, i} \Big(\psi^+_{1,-1}(z) \, \psi^+_{J,M+1}(0) + \psi^+_{J, M+1}(z) \, \psi^+_{1,-1}(0)\Big) \, ,
\ee
where the $\dots$ on the right-hand-side of (\ref{walgebra}) correspond to ambiguities in the addition of such DGS. 

It can be shown~\cite{lee} that the right-hand-side of (\ref{dgs}) (and all other discrete states in the $(-)$ sector) 
can be expressed in terms of products of 
Shur polynomials $S_k (\{-\frac{i}{k\!}\sqrt{2} \partial_z ^k X(0)\})$, with $S_k$ defined through:
\begin{equation}
{\rm exp}\Big(\sum_{k=1}^\infty a_k x^k \Big) = \sum_{k=0}^\infty S_k(\{ a_k\}) \, x^k , \quad \{a_k\} \equiv \{ a_i : \, i \in Z_k \} ~,
\end{equation}
with $S_k=0$ for $k <0$.
The presence of $\partial_z^k X$ terms implies~\cite{iso} that the OP of the DGS operators 
obeys a (classical) $w_\infty$ symmetry algebra of the form (\ref{walgebra})~\cite{lee}:
\be\label{dgsw}
\int \frac{dz}{2\pi \, i} G_{J_1\, M_1}^+ (z) \, G_{J_2\, M_2}^+ (0) = (J_2\, M_1 - J_1, \, M_2) \, G^+_{J_1 + J_2 -1, \, M_1 + M_2 }(0) ~.
\ee
The DGS are the carriers of the conserved $w_{\infty}$ charges. They decouple from the correlation functions of the physical states and can be considered as the symmetry parameters of the theory. 

In the specific context of strings propagating in target-space black-hole backgrounds,  the massless matter particle (``tachyon'') is 
associated with the vertex operator:
\be
             \phi^{c,-c}_{-1/2,0,0} =(g_{++}g_{--})^{-\frac{1}{2}}
F(\frac{1}{2} ; \frac{1}{2} ; 1 ; \frac{g_{+-}g_{-+}}{g_{++}g_{--}}) \, ,
\label{XV}
\ee
where $F$ denotes a hypergeometric function and $g_{ab}$, $a,b =+,-$ represent the components
of a generic SL(2,R) element. The asymptotic form of this operator gives the massless tachyon vertex operator 
of the $c=1$ two-dimensional string theory.

This operator is exactly marginal \emph{only} in the \emph{flat-space} two-dimensional string theory. 
In the presence of a Euclidean black-hole background (or Minkowski, the latter being obtained by 
analytic continuation of the compact `temperature' variable in the cigar metric of \cite{witt}), 
the corresponding  \emph{exactly-marginal} operator is~\cite{Chlyk}
\be
          L^1_0 {\overline L_0}^1 = \phi^{c,-c}_{-1/2,0,0}+ i (\psi ^{++}- \psi^{--}) + \dots \, ,
\label{XVI}
\ee
where
\be
  \psi^{\pm\pm}   \equiv : ({\overline J}^{\pm} )^N (J^{\pm})^N  (g_{\pm\pm})^{j+m-N} \, ,
\label{XVII}
\ee
with
$J^{\pm} \equiv (k-2) (g_{\pm\mp}\partial _z g_{\pm\pm} -
g_{\pm\pm}\partial _z g_{\pm\mp} )$, and
${\overline J}^{\pm} \equiv (k-2) (g_{\mp\pm}\partial _{{\bar z}}
g_{\pm\pm} - g_{\pm\pm}\partial _{{\bar z}} g_{\mp\pm})$,
where $k$ is the Wess-Zumino level parameter \cite{witt}.
The combination $\psi^{++} - \psi^{--}$ generates a level-one
massive string mode, and the dots in equation (\ref{XVI})
represent operators that generate higher-level massive string states.

Another example of an exactly-marginal operator is $L^2_0 {\overline L_0}^2   = \psi^{++} + \psi^{--} + \psi^{-+}
+\psi^{+-} + \dots$, which also involves in an essential way operators for massive string
modes. The coupling corresponding to this world-sheet deformation of the coset model
is associated with a global rescaling of the target space-metric \cite{Chlyk}, and therefore to a global
constant shift of the dilaton field. Thus it produces shifts in the black hole mass~\cite{witt}, 
which is of relevance for the discussion of an evaporating black hole with a time-dependent mass~\footnote{In
the two-dimensional case the horizon has no area, whereas in the four-dimensional case discussed later the 
area, and hence the entropy, is proportional to the mass. The fact that mass and hence entropy are changed by
$L^2_0 {\overline L_0}^2$, which is not an invariant of $W_{1 + \infty}$,
is a consequence of the observation that $W_{1 + \infty}$ symmetry is required for information retention and hence entropy conservation.}.

As discussed in~\cite{emn1,emnQM}, these modes are solitonic, with fixed energy and momentum. 
As such, they are completely \emph{de-localized} in space-time. The vertex operators describing these 
discrete positive-norm states satisfy a $W_{1+\infty}$ symmetry algebra (with the inclusion of a 
vector spin-1 U(1) current), which is the gauge symmetry of the string theory in the black-hole background. 
The conserved $W$-charges are carried by the corresponding discrete gauge states (with zero norm),
whose asymptotic forms coincides with the discrete gauge states of \cite{lee} (\emph{cf.} (\ref{dgs})). 

\subsection{World-Sheet Renormalization-Group Flow, Time and the Increasing Entropy of Evaporating Black Holes} 

Since the flat-space `tachyon' vertex operator (\ref{XV}) is not exactly marginal in a black-hole
background, the corresponding world-sheet renormalization-group (RG) $\beta$ function is non-vanishing.
This induces world-sheet renormalization group flow in the non-critical string theory model of a quantum black hole,
which can be identified with
the entropy increase rate of an evaporating black hole in string theory, as we now review~\cite{emnQM}.

The pertinent world-sheet action has deformations of the form 
\begin{equation}\label{reldeform}
S_\sigma = S_\sigma ^\star + \int _\Sigma d^2 \xi \sqrt{-\gamma} \, g^i \, V_i \, ,
\end{equation}
where $S_\sigma^\star$ is a conformal fixed-point $\sigma$-model action, summation over repeated indices is implied, 
$\gamma$ is a world-sheet metric, 
and $\{ g^i \}$ is an (infinite in general) set of target-space fields associated with the 
corresponding vertex operators $V_i$. 
In a two target-space-dimensional setting, where we start our discussion for instructive purposes~\cite{emn1,emnQM}, 
the only propagating  multiplet consists of massless scalar fields (misleadingly called `tachyons'),
whereas the graviton and (the infinity of) higher-spin multiplets are topological ``massive'' states with discrete momenta. 
As we discussed in \cite{emn2} and review later in this article, such topological states exist also in higher-dimensional 
target space-times, so their presence is rather generic. 

Introducing a global world-sheet RG scale ${\mathcal \mu}$,
and defining ${\mathcal T} \equiv {\rm ln}\, \mu$, we consider 
renormalised world-sheet couplings $g^i$, `running' with ${\mathcal T}$, according to the following world-sheet 
RG equation:
\be\label{beta}
\frac{d\, g^i}{d\, {\mathcal T}}  \equiv \beta^i \, ,
\ee
where $\beta_i$ is the RG $\beta$ function for the `coupling' $g^i$ of the two-dimensional world-sheet field theory~\footnote{More 
strictly speaking, in string theory the target-space dependences of the `couplings' $g^i$ imply
some diffeomorphism variations, which lead to the replacement of the corresponding
RG functions $\beta^i$ by the corresponding Weyl anomaly coefficients, but such complications are not 
relevant for our main arguments below, so we omit them here. For details see \cite{emn1}.}.

The presence of relevant deformations in the world-sheet $\sigma$ model, 
when only the propagating modes of the string multiplet are taken into account, 
calls for Liouville dressing~\cite{liouville}. This requires introducing an extra $\sigma$-model field, 
the Liouville mode $\rho$ that, for supercritical deformations such as those in the black hole case of interest~\cite{emn2}, 
has a target time-like signature, and may be identified with target time, flowing irreversibly~\cite{emnQM}. 
The Liouville mode can be viewed as a local RG scale, as required because
the string world sheet is generally curved. For a non-marginal coupling $g^i$,
e.g., a massless `tachyon' field in the target-space of the two-dimensional field theory in the vicinity of a 
spherically-symmetric stringy black hole, the Liouville RG equation replacing (\ref{beta}) is given by~\cite{liouville}:
\be\label{liouvrg}
{\ddot g}^i + Q(\rho_0) \, {\dot g}^i = -\beta^i(g) = -{\mathcal G}^{ij}\frac{\delta V[g]}{\delta g^j}~, \quad 
{\dot A} \equiv \frac{d}{d\, \rho_0}\, A~,
\ee
where the overdot denotes a derivative with respect the Liouville world-sheet zero mode $\rho_0$ 
(with a canonically-normalised term in the world-sheet action~\cite{liouville}), 
$\beta$ is the  world-sheet RG function for the (relevant) coupling $g^i$, given by (\ref{beta}),
expressed as a gradient of an effective target-space potential for the fields/couplings $\{g^j\}$, 
and ${\mathcal G}_{ij}$ is the Zamolodchikov metric in the space of string theory 
models~\cite{emnQM}. The quantity $Q(\rho_0)$ is the square root of the central charge deficit $Q^2$,
which is $> 0$ for supercritical $\sigma$ models~\cite{aben,emnQM}. 

Upon identifying (see~\cite{emnQM}) the flow of the world-sheet zero-mode of the Liouville field with the opposite flow of the 
RG scale ${\mathcal T}$ and of the target time (for the supercritical string case,  as explained in detail in ref.~\cite{emnQM}):
\be\label{time}
t = - {\mathcal T} (= - \rho_0) \, ,
\ee
and taking into account~\cite{aben} that in the case of bosonic target-space background fields $g^i$ 
there are `tachyonic' mass shifts in their dispersion relations, i.e., terms in the potential:
\be\label{potential}
V[g]_{\rm boson} \ni - \frac{1}{2}\, Q^2 g^i \, g^j \delta_{ij} ~,
\ee
one may move half of this term to the left-hand-side of the Liouville RG equation (\ref{liouvrg}). 
One may then write the Liouville flow as an equation of motion for the fields $g^i$ obtained from an one-dimensional 
gauge theory with time (and only) component of the gauge potential $2A_0 = i Q(t)$: 
\be\label{cov}
{\mathcal L}_{\rm Liouv} = (D_t g^i)^\dagger {\mathcal G}_{ij} D_t g^j - {\tilde V}[g]~, \quad D_t = \partial_t -iA_0(t) ~,
\ee
where $\dagger$ denotes hermitian conjugation, and ${\tilde V}$ contains the remaining part of the 
tachyonic mass shift terms of the form $-\frac{1}{4} Q^2 g^i g^j \delta_{ij}$ along with 
the rest of the interaction terms for the fields $g^i$.  
We shall make use of (\ref{cov}) when we discuss scalar modes of Hawking radiation in the horizon of a spherically-symmetric black hole. 

We also remark that, upon the identification (\ref{time}) in a non-critical string theory~\cite{emnQM}, the effective central charge 
$C[g]=Q^2 $ obeys an irrerversible flow equation:
\be\label{increase}
 \frac{d}{d \rho_0} Q^2 = - \frac{d}{dt} Q^2 \sim -\beta^i {\mathcal G}_{ij} \, \beta~^j  < 0~,
 \ee
provided the Zamolodchikov metric is positive definite. This happens in string theories with constant dilaton backrounds, 
and in Euclidean target space-times, as used to describe a finite-temperature black hole space-time.

Given that the central charge counts the degrees of freedom of a system, 
the relation (\ref{increase}) implies that the system flows towards an increase of its degrees of freedom 
(and thus its entropy) as time $t$ progresses, or equivalently during the evolution from infrared to ultraviolet on the world sheet, 
whenever a string propagates in a non-conformal background. Hence, the entropy associated with massless `matter' increases 
inexorably, {i.e}, information is lost, \emph{if the higher-level string modes in (\ref{XVI}) are neglected.} 
Conversely, if the discrete solitonic string modes (\ref{XVI}), (\ref{XVII}) are
taken into account, the corresponding RG $\beta$ function vanishes, and entropy does not increase
with the world-sheet RG flow, which we identify with the target-space temporal time flow in our approach.
Thus, there is no information loss: it is stored by the higher-level string modes.

These topological modes are not detectable in a local scattering experiment, 
leading to an apparent `loss' of quantum coherence, which is an artefact of the phenomenological
truncation of the scattering process within a local effective field theory (LEFT) framework. 
Associated with this apparent `loss' of quantum coherence there is an apparent `increase'
in entropy at a rate quantified by the right-hand-side of (\ref{increase}), since the \emph{truncated} RG 
$\beta^i$ functions of the non-marginal propagating modes do not vanish. 
Nevertheless, the conserved W-hair charges are in principle measurable, 
and ways for doing so in principle have been outlined in \cite{emn2}. 

\subsection{Embedding of the Two-Dimensional Black Hole in Four Space-Time Dimensions}

The coset singularity structure of the two-dimensional stringy black hole and
generic properties of its associated discrete states have counterparts
for spherically-symmetric black-hole configurations in four space-time dimensions.
This can be seen by embedding the two-dimensional coset describing the singularity 
in a four-dimensional space-time~\cite{emn2} with the structure SU(1,1)/U(1) $\otimes ~ S^{2}$, 
where $S^2$ is a two-dimensional manifold with the topology of the sphere that is to be identified with the horizon 
of the four-dimensional black hole. This enables us to place our arguments on the importance of 
$W_\infty$ symmetries in a more generic perspective. 

To see this, consider a spherically-symmetric gravitational
background of black-hole type, which is a solution of the generalised Einstein equations in the effective
field theory derived from string theory. The metric tensor takes the form
\begin{equation}
ds^2 =g_{\alpha\beta} dx^{\alpha}dx^{\beta} + e^{W(r,t)}d\Omega^2 \, ,
\label{sphere}
\end{equation}
where $W(r,t)$ is a non-singular function, the $r,t$ coordinates are denoted by $x^{\alpha,\beta}$,
and the line element on a fixed spherical surface is denoted by $d\Omega ^2 = d\theta ^2 + \sin^2 \theta d \phi ^2 $.
The standard Schwarzschild solution describing a spherically-symmetric four-dimensional
black hole can be cast in the form (\ref{sphere}) with an appropriate transformation of variables.

When written in Kruskal-Szekeres coordinates, the Schwarzschild solution takes the form~\cite{thorn}:
\begin{equation}
ds^2 = -\frac{32M^3}{r}e^{-\frac{r}{2M}}du dv + r^2 d\Omega ^2 \, ,
\label{krusk}
\end{equation}
where $r$ is a function of $u,v,$ given by
\begin{equation}
  (\frac{r}{2M} - 1) e^{\frac{r}{2M}}=-uv \, .
\label{reln}
\end{equation}
Although the two-dimensional metric components depend
on the variables $u,v$, the black hole solution is nevertheless static~\footnote{Moreover, 
in pure gravity all the {\it classical} spherically-symmetric solutions to the equations of
motion obtained from higher-derivative gravitational actions with an arbitrary number of curvature tensors
are {\it static}~\cite{whitt}, and a similar result holds for stringy black holes at tree level.}.
Changing variables to
\begin{eqnarray}
\nonumber
e^{-\frac{r}{4M}}u=u' \, , \\
e^{-\frac{r}{4M}}v=v'
\label{change}
\end{eqnarray}
we can write the two-dimensional metric in the form
\begin{equation}
 g_{bh}(u',v')= \frac{e^{D(u',v')} du'dv'}{1-u'v'} \, ,
\label{bh}
\end{equation}
with the scale factor being given by $ 16M^2 e^{-\frac{r'(u',v')}{2M}}J(u',v')$,
where $r'$ is the coordinate $r$ re-expressed in therms of the coordinates $u',v'$, and
$J$ is the Jacobian of the transformation of the area element $dudv$.
The metric (\ref{bh}) is a conformally-rescaled form of Witten's two-dimensional  black hole solution~\cite{witt}. The
latter is described by an exact conformal field theory, so the same is true after this conformal rescaling,
which simply represents a change of renormalisation scheme
according to the $\sigma$-model point of view. The 
function $D(u,v)$ in (\ref{bh}) can also be regarded as part of the two-dimensional dilaton in the given renormalisation
scheme.  

The global properties, such as singularities, are the same as in the two-dimensional string case.
In particular, the infinite-dimensional W-symmetry associated with the
SU(1,1)/U(1) coset structure of the dilaton-graviton sector in the two-dimensional model
is also a model-independent feature of spherically-symmetric
four-dimensional string configurations. Such structures are associated with 
topological solitonic non-propagating states, which are
spherically-symmetric solutions of the low-energy equations of motion obtained from string theory in a manifold
with topology SU(1,1)/U(1) $\times {\mathcal M}^2$,
where ${\mathcal M}^2$ is a two-dimensional manifold of constant curvature.
They correspond to jumps in the number of degrees of freedom at discrete energy-momentum values,
resulting from the relaxation of certain gauge theory constraints, as shown below. 
The simplest example of such a manifold is where
${\mathcal M}^2=S^2$, which describes a spherically-symmetric four-dimensional black hole.
The infinite number of associated discrete topological (non-propagating) states, with discrete energy-momentum values,
couple to the massless propagating `tachyon' string matter and render the 
associated $\sigma$-model action conformally invariant, as in the two-dimensional stringy black hole~\cite{witt}, described above.  

A $w_\infty$ symmetry also arises in the phase space of matter
coupled to another example of a two-dimensional string theory embedding in four dimensions~\cite{zanon}, namely a 
four-dimensional extremal solitonic black hole background in the context of 
$N=2$, $D=4$ supergravity.
This is a BPS solution that interpolates between an AdS$_2 \times H^2$
geometry (where AdS$_2$ refers to the radial-coordinate/time
part of the space-time and $H^2$ is is a hyperbolic two-dimensional manifold of constant curvature
describing the angular part of the space-time)
that characterises the space-time near the horizon of the black hole
and a maximally-supersymmetric AdS$_4$
space-time at large radial distances. It was shown in~\cite{zanon} 
showed that a quantum-mechanical massive particle with non-trivial magnetic charge
in the near-horizon geometry has dynamics described by a one-spatial-dimensional Hamiltonian $H$, 
with a $w_\infty$ symmetry that preserves the two-dimensional phase-space area symplectic form
$\Omega  = dp \wedge dq - dH \wedge dt $, where $q$ is the spatial coordinate, $p$ is the canonical momentum and $t$ is the time. 

The energy spectrum of this particle is continuous and bounded from below: $E>0$,
but  the ground state is non-normalisable, with an infrared (IR) divergence,
which was regularised in \cite{zanon} by putting the system in a box.
The IR-regularised system is also invariant under a $w_\infty$ that contains a Virasoro symmetry (\ref{Vira}),
which can be associated 
with the asymptotic symmetries of the AdS$_2$ space time,
{i.e.}, the diffeomorphisms that leave invariant the AdS$_2$ metric, whose
quantum version includes a central extension. 
Such asymptotic symmetries are symmetries of the quantum-gravity scattering matrix for the full 
four-dimensional AdS$_2 \times H^2$ extremal black hole of \cite{zanon}~\footnote{An asymptotic 
symmetry of the quantum-gravity scattering matrix under supertranslations of generic black hole 
backgrounds has been examined in \cite{strominger,strominger2}.}.

The particle system in this example is characterised by an infinite set of conserved charges 
corresponding to diffeomorphisms 
preserving a two-dimensional symplectic area two-form $\Omega$ defined for the ``coordinates'' $x, y$:
\be\label{sympl}
\Omega = dy \wedge dx \, .
\ee
These area-preserving diffeomorphisms are generated by the quantities
\be\label{gen}
v_m^\ell = y^{\ell + 1} \, x ^{\ell + m + 1} \, ,
\ee
where $\ell$ and $m$ are integers. The Poisson brackets of these generators 
satisfy the classical $w_\infty$ algebra
\be\label{winfty}
\{ v_m^\ell, \, v_{m^\prime}^{\ell^\prime} \} = [m \, (\ell^\prime + 1) - m^\prime \, ( \ell + 1) ]\, v_{m + m^\prime}^{\ell + \ell^\prime} \, .
\ee
This includes a Virasoro symmetry generated by the operators $L_n = v_{n}^{0}$,
whose Poisson brackets obey the algebra
\be
\{ L_n, \, L_m \} =  (m-n)\, L_{m+n}~,
\label{Vira}
\ee
which is a subalgebra of the $w_\infty$ algebra (\ref{winfty}). 

In the case of the effective two-dimensional mechanics of particles in the near-horizon geometry of the four-dimensional black hole, 
the roles of the symplectic coordinates $x,y$ are played by appropriate combinations of the \emph{phase-space}
coordinates of the particle~\cite{zanon}, and hence the system is completely integrable.
The role of the $w_\infty$ algebra for the particle near the horizon of the four-dimensional black hole 
of \cite{zanon} is exactly analogous to that 
preserving the phase-space area for massless `tachyonic' string matter 
in the two-dimensional stringy black hole~\cite{emn1} - or its four-dimensional extension with topology 
SU(1,1)/U(1) $\times S^2$ - as discussed previously.

It was suggested in \cite{emn1} that the infinite set of $W$ charges in the quantum version of the classical $w_\infty$ symmetry
provide an \emph{infinite set of discrete gauge hair} (called W-hair), which
maintains the quantum coherence for the two-dimensional stringy black hole. This was based on the fact
that the quantum-gravity scattering matrix obtained from correlation functions of marginal world-sheet vertex operators is invariant under the
quantum $W$ symmetry.
The existence of an infinite set of conservation laws for a 
particle in the near-horizon geometry of the black hole discussed in~\cite{zanon}, and hence an infinite set of conserved charges $v_m^\ell$, 
also guarantees quantum coherence by retaining information during the black-hole evaporation.  
The elevation of the classical phase-space area-preserving $w_\infty$ symmetry algebra to a
quantum algebra capable of preserving coherence necessarily involves the discrete massive topological states of the string, as discussed above.
It is their mixing with the propagating massless matter
states that guarantees the conformal symmetry of the corresponding vertex operators in a stringy black hole background~\cite{emn1}, 
and thereby preserves quantum coherence.

There are infinitely many discrete topological gauge states in a string theory in a $D$-dimensional target space,
which have a similar nature to those in the two-dimensional case~\cite{pol,klebanov,lee} that appear
in (\ref{dgs}). The existence of these states can be understood by examining
the gauge conditions for a rank-$n$ tensor multiplet:
\begin{equation}
   D^{\mu_{1}}A_{\mu_{1}\mu_2...\mu_n}=0 \, ,
\label{gauge}
\end{equation}
where $D_{\mu}$ is a (curved-space) covariant derivative.
To see this,
consider for example weak gravitational perturbations around
flat space with a linear dilaton field
of the form $\Phi(X)=Q_{\mu}X^{\mu}$. In this case one may
Fourier transform (\ref{gauge}) to find
\begin{equation}
    (p + Q)^{\mu_1} {\tilde A}(k)_{\mu_1\mu_2....\mu_n}=0 \, .
\label{gauge2}
\end{equation}
This shows that the number of degrees of freedom increases at the discrete momentum $p=-Q$.
Since this momentum is fixed, it corresponds to a complete delocalised state,
which should be regarded as a quasi-topological, non-propagating soliton-like state. Such states
carry a small statistical weight in ordinary string theories, relative to the continuous
spectrum of the continuum string modes. However, these discrete states
assume particular importance when strings propagate in spherically-symmetric four-dimensional
background space-times. These backgrounds are effectively
two-dimensional, and Ward identities of the form (\ref{gauge}) may be sued to gauge away all the
transverse modes of higher-rank tensors, except for
these topological modes. These $s$-wave
topological modes constitute the final stages
of the evaporation of four-dimensional
spherically-symmetric black holes~\cite{emn1}, and play key roles in maintaining quantum coherence \cite{emn1,emn2}. 

These discrete solitonic states can be regarded
as singular gauge configurations~\cite{wmeasure}, similar to the discrete gauge states in two-dimensional strings (\ref{dgs}), 
and their conserved $W$-charges could in principle be
measured via generalized Aharonov-Bohm phase
effects. These higher-spin topological string states also leave their imprints via
selection rules in the scattering matrix, where they appear as (resonance) poles at discrete energies and momenta. 
There is an infinite set of such black-hole soliton states in the stringy black-hole case, which can be classified by the quadratic
Casimir and `magnetic' quantum numbers of an internal symmetry group~\cite{wmeasure}.
These resonances appear at calculable energies and decay into
distinctive combinations of light final-state particles. 

The stringy scattering matrix in such a black-hole background is well defined in general,
since the correlation functions among the appropriate exactly-marginal vertex operators on the world-sheet are unitary,
since these operators contain an infinity of non-propagating topological states as well as the parts 
corresponding to propagating string states. In practice, 
these delocalised states cannot be detected in laboratory scattering experiments, 
since they involve a finite number of localised (in space-time) particle states.
Hence, there would apparently be decoherence from the point of view of a local low-energy observer,  
even though there are no pathologies in the full stringy theory of quantum gravity.

\subsection{Horizon-Area-Preserving $w_\infty$ Classical Symmetries}

The area of a black hole be be regarded classically as a conserved Noether charge~\cite{wald}. 
The diffeomorphisms 
on the black-hole horizon that preserve the horizon area belong to the classical $w_\infty$ symmetry algebra of
transformations of the horizon coordinates. These preserve the area of the 
horizon of an isolated spherically-symmetric four-dimensional
black hole by construction, thereby conserving its entropy in agreement with the results of \cite{wald}. 

A classical $w_\infty$ symmetry may manifest itself in a number of ways, as we now discuss. 
The horizon of a stringy black hole
may be regarded as a thick brane, which is known to be describable as an SU($\infty$) gauge theory~\cite{fit,ilio},
leading to the infinitely-coloured SU($\infty$) black hole model of~ \cite{winstanley}. In this approach,
open-string states terminating on the horizon brane carry the SU($\infty$) charges. 
Such an SU($\infty$) symmetry is classically isomorphic to the classical $w_\infty$ algebra preserving a two-dimensional area, 
which can be identified in this case with the area of the horizon of the infinitely-coloured, spherically-symmetric black hole. 

As we have discussed, there is a classical $w_\infty$ algebra that
preserves the two-dimensional area of an `internal space' with the topology of a 
sphere~\cite{iso,wcontr}. The question is then whether this `internal' sphere can be identified with the 
horizon of the four-dimensional spherically-symmetric Schwarzschild black hole.
In order to analyse this issue, we considered in~\cite{emn2015} examples of four-dimensional spherically-symmetric 
black holes with infinitely-coloured hair, which realise explicitly 
a classical $w_\infty$ symmetry as discussed in the previous paragraph.  
Thes black-hole solution of SU($N\to \infty$) gauge theory is formulated in a 
four-dimensional AdS space-time with negative cosmological constant, 
which plays the role of a regulator making the black-hole solution well-defined~\cite{winstanley}. 
This anti-de-Sitter (AdS) regulator was given physical significance via the AdS/CFT bulk/boundary correspondence, 
and turned out to be physically important, as we argued in \cite{emn2015} and discuss briefly below.

Our interest in these black holes is motivated by the
classical isomorphism between SU($N \to \infty$) and 
$w_\infty$~\cite{iso,fit}~\footnote{The geometry of SU(N) gauge theories with finite $N$
is non-commutative~\cite{ilio}, commutativity appearing only in the limit $N \to \infty$.}.
To see this correspondence, one replaces the SU(2) generators $S_i$ by rescaled versions,  $T_i \equiv \frac{2}{N}\, S_i$,
and finds~\cite{emn2015} that in the limit $N \to \infty$ the $N^2 -1$ matrices $T_{\ell\, m}^{(N)}$ obey:
\be\label{compoisson} 
\frac{N}{2i} \, \Big[ T_{\ell, \, m}^{(N)}\, , \, T_{\ell^\prime, \, m^\prime}^{(N)}  \Big]  \rightarrow \,  \quad \{ Y_{\ell, \, m} \, , \, Y_{\ell^\prime, \, m^\prime} \}, \quad N \to \infty \, ~,
\ee
where we denote by $Y_{\ell\, m}(\theta, \phi)$ the spherical harmonics on the sphere $S^2$.
It is well-known that the (classical) Poisson algebra of the spherical harmonics is that of
$SDiff(S^2)$, the infinite set of area-preserving diffeomorphisms on the sphere, :
\bea\label{sdiffs}
\{ Y_{\ell, \, m} \, , \, Y_{\ell^\prime, \, m^\prime} \} &=&\frac{M(\ell + \ell^\prime -1, \, m + m^\prime)}{M(\ell, \, m)\, M(\ell^\prime, \, m^\prime)}\, (\ell^\prime \, m - \ell \, m^\prime) \, Y_{\ell + \ell^\prime -1, \, m + m^\prime}  \nonumber \\
&+ & \sum_{n=1} \, g_{2n}(\ell, \, \ell^\prime)\, C^{\ell + \ell^\prime -1 -2n, \, m + m^\prime}_{\ell, \, m, \, \ell^\prime, \, m^\prime} \, Y_{\ell + \ell^\prime -1-2n, \, m + m^\prime}~,
\eea
where the the structure constants $C$ are given in the fourth paper in~\cite{iso},
and $M $ and $g_{2n}$  are normalisation factors. 
This algebra  is isomorphic to the classical area-preserving $w_\infty$ algebra.

Writing the gauge fields of the SU($N \to \infty$) gauge theory using the matrices $T_{\ell, \, m}^{(N)}$ as a basis, 
obeying (\ref{compoisson}), we see that this area-preserving diffeomorphism symmetry preserves the horizon area
in this example of an infinitely-coloured gauge black hole, upon identification of the `internal' sphere $S^2$ with the 
horizon sphere of the spherically-symmetric SU($\infty$) black hole. In this case, the entropy of the black hole
can be preserved classically by the $w_\infty$ symmetry and its associated infinite set of $W$ hair. 

Viewing this SU($\infty$) gauge model as 
a low-energy limit of some string-theory black hole indicates, according to our world-sheet RG interpretation of the target time,
that the classical conservation of the horizon area reflects the 
conformal invariance of the corresponding world-sheet field theory, 
which guarantees that the right-hand-side of (\ref{increase}) vanishes because the 
$\beta^i$ functions of relevant combinations 
of the couplings $g^i$ are zero. Here the set $\{ g^i \}$ comprises the 
graviton $G_{\mu\nu}$ and the SU($\infty$) gauge field
modes: $A_\mu^a, a=1 \dots \infty$. In \cite{emn2015} the AdS regulator of the SU($\infty$) model has been given 
physical significance in guaranteeing the vanishing of 
these $\beta$ functions, upon mixing the graviton with the gauge-field  contributions, analogously to our
previous stringy black-hole case,
where higher-spin states mix with the massless string matter contributions to ensure conformal invariance on the world-sheet.

Proceeding further by describing the horizon of the four-dimensional black hole as a two-brane~\cite{emn2015},  
one can treat the recoil when an open string, representing a matter state, 
meets the horizon surface. When one of the ends of the open string attaches to the horizon, 
the latter recoils so as to conserve momentum. 
This horizon recoil causes it to fluctuate in a way characterised by a logarithmic 
conformal field theory on the world-sheet~\cite{kogan, EllisDbrane} that
describes the transfer of information~\cite{mislays}. Such a recoiling black-hole horizon may be represented 
as a `thick' stack of $N \to \infty$ concentric D-branes. 
In the case of a macroscopic black hole, whose horizon is large compared to the wavelength of the infalling matter, 
these concentric branes may be regarded approximately as a stack of parallel and flat $N \to \infty$ branes,
which are equivalent to an SU($N \to \infty)$ gauge theory~\cite{maldacena}.
This can be seen intuitively by considering the different ways ($N^2-1$ in the case of an SU(N) gauge theory) 
in which an open string can be attached to a stack of $N$ D-branes. Therefore, when matter 
reaches the thick brane model of the black hole horizon, the recoil is described by excitations that
carry the SU($\infty$) charges. These correspond to the infinite $W$ hair of the black hole and
the classical horizon area-preserving $w_\infty$ symmetry discussed above.
In this example the SU($\infty$) symmetry plays a role as a coherence-preserving symmetry of the 
scattering matrix involving the SU($\infty$) black hole. 

However, the presence of classical infinite-dimensional horizon-preserving symmetries is more generic than the above example, 
and is in fact associated with the observation of~\cite{bonora} that the entire spectrum of Hawking radiation of black holes (either
Schwarzschild as we consider here or rotating Kerr type) may be represented in terms of higher-spin currents of the 
associated conformal field theory near the horizon surface. We proceed to discuss this case
next, and then place it in the context of our string theory considerations. 

\section{Hawking Radiation from Generic Schwarzschild Black Holes and $W_{1+\infty}$ Algebras \label{sec:hr}}

In order to discuss explicitly the connection of Hawking radiation to our $W_{1+\infty}$ symmetries,
we first review briefly some interesting results~\cite{bonora,higherspin} connecting Hawking radiation in
rather generic, non-stringy, spherically-symmetric black holes to a $W_{1+\infty}$ algebra realised by 
higher-spin states associated with the moments of the Hawking radiation. These $W_{1 + \infty}$ currents are 
sourced by background fields of these higher-spin states, which can be identified with the discrete gauge states
in our stringy approach. 

The effective two-dimensional conformal field theory representation~\cite{wilczek} of
the dynamics of matter fields in the near-horizon geometry of a spherically-symmetric black hole~\cite{emn1}
is crucial for the connection of Hawking radiation to $W_\infty$ algebras. It is known that  
the outgoing Hawking quanta radiated from the horizon of a spherically-symmetric black hole break general covariance. 
As shown in \cite{wilczek}, this symmetry is restored through the cancellation of the corresponding gravitational 
anomalies in the quantum-gravity path integral for a (1 + 1)-dimensional black body at the Hawking temperature of a 
black hole~\cite{hawk}. One can represent the effective two-dimensional field theory of the Hawking 
radiation at the black hole horizon as a two-dimensional field theory with an infinity of two-dimensional 
conformal quantum fields obeying a thermal spectrum, with the left-movers corresponding to infalling matter and the
right-movers to outgoing matter. 

We restrict ourselves here to Schwarzschild black holes, which emit Hawking radiation with a Planck distribution 
\be\label{planckdistr}
N^\pm (\omega) = \frac{1}{e^{\beta \, \omega} \pm 1}
\ee
where $\omega$ is the frequency (energy) of the radiated quantum, $\beta$ is the inverse of the Hawking temperature~\cite{hawk}, 
and + (-) for fermions (bosons) respectively. The full spectrum of the radiation is encoded in the infinite set of
moments of the Hawking radiation spectrum~\cite{higherspin}:
\be\label{highermomplus}
F_{2n}^+ = \int_0^\infty \frac{d\omega}{2\pi} \omega^{n-1} \, N^+ (\omega) = (1 - 2^{1-2n})\, \frac{B_{2n}}{8\pi \, n}\, \kappa^{2n}~,
\ee
or
\be\label{highermomminus}
F_{2n}^- = \int_0^\infty \frac{d\omega}{2\pi} \omega^{n-1} \, N^- (\omega) =  \frac{B_{2n}}{8\pi \, n}\, \kappa^{2n}~,
\ee
where the $B_{2n}$ are the Bernoulli numbers, and $\kappa = 2\pi/\beta $ is the surface gravity of the black hole. 
For example, the energy flux is given by the 
second moment  of $N^\pm (\omega)$, $F_2 (\omega) = \int_0^\infty \frac{d\omega}{2\pi} \omega \, N^\pm (\omega) $. 

It was proposed in~\cite{higherspin} that, generalising the connection of the energy flux of the black hole to a spin-two current
with matrix element $F_2 (\omega)$,  the higher moments $F_{2n}$, $n > 1$ could be identified as
matrix elements of phenomenological higher-spin currents that could be regarded as generalisations of the energy-momentum tensor. 

These higher-spin currents can be expressed~\cite{higherspin} as (appropriately normal-ordered)
products of two-dimensional boson and fermion fields 
and their space-time derivatives. In terms of the light-cone variables 
\be\label{lightcone}
u= t + r_\star, \, \quad v=t - r_\star~, \quad r_\star : \, \frac{\partial r_\star}{\partial r} = f(r)^{-1} ~, 
\ee
where $r_\star$ is the so-called `tortoise' coordinate, a metric of the form: $ds^2=f(r) dt^2 - \frac{1}{f(r)} dr^2$, 
which may be used to represent the effective two-dimensional space-time in the near-horizon geometry of the spherically-symmetric 
black hole, becomes that of a conformally-flat space-time:
\be\label{nhg}
ds^2 = e^{2\varphi (t,r^\star)} \Big(dt^2 - d(r^\star)^2 \Big) = e^{2\varphi (u, v)}\, du \, dv ~, \,\quad  e^{2\varphi(u, \, v)} = f(r)~.
\ee
Outgoing scalar radiation is then described by holomorphic ($v$-independent) currents of the form~\cite{higherspin}:
\be\label{holomorphic}
{\rm Outgoing~radiation~scalar~currents:} \;  J^B_{uu\dots u} = {\rm linear~combinations~of} \, : (-1)^{n + m} \, \partial_{u}^m \, \phi \, \partial_u^{2n - m} \, \phi :~,
\ee
and for fermions $\psi (u)$ one has:
\be\label{holomorphicfermions}
{\rm Outgoing~radiation~fermion~currents}: \;  J^F_{uu\dots u} = {\rm linear~combinations~of} \, : \overline \psi \, \partial_u^{n}\, \psi :~.
\ee
where $: \dots :$ denotes the appropriate normal ordering, as defined in \cite{higherspin}. 

When representing the higher moments of the Hawking radiation in terms of conformal fields on the horizon, 
there are ambiguities in the relative coefficients of the various terms appearing in the holomorphic currents (\ref{holomorphic}) and (\ref{holomorphicfermions}), and the currents are not normalisable in general.
These issues were resolved in~\cite{bonora} by requiring that the coefficients be fixed by a symmetry principle, specifically 
by postulating that there is a a $W_\infty$ algebra on the horizon of the black hole, generalising the Virasoro algebra~\footnote{Their higher-derivative holomorphic structure, $\partial^n_z \chi $ ($\chi=\phi, \psi$)~\cite{iso} led one to 
expect that the currents could realise such an infinite-dimensional algebra.}.

In the case of flat two-dimensional space-times, upon Euclideanisation 
and replacing the coordinates $u, v$ by the complex variables $z$, $\overline z$ respectively, 
the bosonic $w_\infty$ currents with conformal spin $s$ can be written as 
follows~\cite{bonora}~\footnote{From now on we use complex scalar fields to represent the Hawking spectrum, 
so as to make a direct correspondence with the $W_{1+\infty}$-algebra formalism.}:
\be\label{wbosonic}
j_{z\dots z}^{(s) B} = q^{s-2}\frac{2^{s-3}\, s\, !}{(2s-3)!\,!}\, \sum_{k=1}^{s-1} \, (-1)^k \, \Big[\frac{1}{s-1}\, \begin{pmatrix} s-1 \\ k \end{pmatrix} \, \begin{pmatrix} s-1\\ s-k \end{pmatrix} \, \Big] \, : \, \partial_z^k \phi(z) \, \partial_z^{s-k} \overline{\phi}(z) \, :
\ee
where $q$ is a complex deformation parameter~\cite{iso}.
The holomorphic free fields $\phi(z)$ are assumed to have the following non-vanishing two-point function  
$\langle \phi(z) \, {\overline \phi}(z^\prime) \rangle = -{\rm ln} (z-z^\prime)$.
The spin $s=2$ current is independent of the deformation parameter $q$, as expected because
this current can be unambiguously identified with the holomorphic stress tensor
$$ j_{uu}^{(2)} = -2\pi \, T^{\rm hol}_{uu}~,$$
but the higher-spin currents depend on parameter $q$.
It can be fixed~\cite{bonora} by demanding that the currents (\ref{wbosonic}), 
when {covariantised} as appropriate in the curved space-time of the spherical symmetric black hole,
reproduce the higher moments of the Hawking flux, leading to $q=-i/4$~\cite{bonora}. 

The work of \cite{bonora} also discussed appropriately-normalised higher-spin currents for the fermion fields,
working in a Kerr metric with Kerr parameter $a \ne 0$, 
corresponding to a non-trivial angular momentum of the rotating black hole. 
In this case, the corresponding gravitational covariant derivative for Dirac fermions in the effective two-dimensional 
space-time (\ref{nhg}) has an extra gauge U(1) potential, proportional to the 
parameter $a$. The fermionic currents then
satisfy a $W_{1+\infty}$ algebra, as a result of the inclusion of the conformal spin-one, 
U(1) gauge vector current in the construction~\cite{bonora}.  One may take the limit $a \to 0$ to 
consider the Schwarzschild black-hole case
and, although the background gauge potential in such a case vanishes, 
nevertheless the corresponding current is non-zero, which implies that the U(1) spin-one current has to be included in the construction
in the case of fermion matter. For completeness, we note that in a Euclideanised flat space-time with 
coordinates $z, \overline z$, these currents are given by~\cite{iso}:
\be\label{wfermionic}
j_{z\dots z}^{(s) F} = q^{s-2}\frac{2^{s-3}\, s\, !}{(2s-3)!\,!}\, \sum_{k=1}^{s-1} \, (-1)^k \, \frac{1}{s}\, \begin{pmatrix} s-1\\ s-k \end{pmatrix}^2 \,  \, : \, \partial_z^{s-k} \Psi^\dagger (z) \, \partial_z^{k-1} \overline{\Psi}(z) \, : \, ,
\ee
where $\Psi (z)$ denotes the two-dimensional conformal fermion field, and $: \dots :$ again denotes the appropriate normal ordering. 

The fact that the currents $j^{(s)}$ are quadratic in the fields is an important feature that we explore in the next Section, when we 
discuss the gauging of the $W$-algebra and its r\^ole in providing a mechanism for the classical conservation of the horizon area 
of a black hole. This feature is carried over in the covariantised currents, which are at most linear functions of the 
currents $j^{(s)}$ and their gravitationally-covariant derivatives $\nabla j^{(s)}$, as can be explicitly seen 
by the expressions of the first few of them, that we list below for concreteness~\cite{bonora}. 
The bosonic currents with conformal spin $\le 6$ are~\cite{bonora} 
\begin{eqnarray}\label{Jsjs}
	J^{(2)B}_{uu} &=& j^{(2)B}_{uu}  - \frac{ \hbar }{6}  \ET  \, , \nn
	J^{(3)B}_{uuu} &=&  j^{(3)B}_{uuu}  \, , \nn
	J^{(4)B}_{uuuu} &=&  j^{(4)B}_{uuuu}  + \frac{ \hbar}{30}   \ET^2 +
\frac{2}{5} \ET J^{(2)B}_{uu}   \, , \nn
	J^{(5)B}_{uuuuu} &=&  j^{(5)B}_{uuuuu} + \frac{10}{7}   \ET  J^{(3)B}_{uuu}  \, , \nn
	J^{(6)B}_{uuuuuu} &=&  \left(
-\frac{2 \hbar }{63  }\ET^3
+\frac{5 \hbar }{504  }\left(\partial _u\ET\right)^2
-\frac{\hbar }{126  }\ET\partial _u^2\ET \right. 
 \nn
&&
-\frac{2}{3}\ET^2J^{(2)B}_{uu}
-\frac{1}{21}\ET\nabla _u^2J^{(2)B}_{uu}
-\frac{1}{21}\left(\partial _u^2\ET\right)J^{(2)B}_{uu}
+\frac{5}{42}\left(\partial _u\ET\right)\nabla _uJ^{(2)B}_{uu}
\nn
&&
\left.
-\frac{5}{21}\Gamma \ET\nabla _uJ^{(2)B}_{uu}
-\frac{5}{21}\Gamma ^2\ET J^{(2)B}_{uu}
+\frac{5}{21}\Gamma \left(\partial _u\ET\right)J^{(2)B}_{uu}\right) 
-\frac{5}{24}\ET J^{(4)B}_{uuuu}
+j^{(6)B}_{uuuuuu} \, , 
\end{eqnarray}
where
\begin{equation} 
\Gamma = \partial_u \varphi~, \qquad 
\ET = \partial^2_u \varphi -  \frac{1}{2} 
\left( \partial_u \varphi \right)^2 =  \partial^2_u \varphi -  \frac{1}{2} \, \Gamma^2~,
\end{equation}
where $\varphi (u,v)$ was defined in (\ref{nhg}), and is associated with the effective two-dimensional 
curved metric near the horizon written in a conformally-flat form. 

The corresponding fermionic currents with spin $\le 5$ are:
\bea\label{kerrcurrent}
J^{(1)F}_{u} &=& j^{(1)F}_{u} + \frac{i\hbar}{2q} A_{u}  \, , \nn
J^{(2)F}_{uu} &=& \left(-\frac{T}{12}\right) \hbar -
2 A_u J^{(1)F}_{u}+j^{(2)F}_{uu} \, , \nn
J^{(3)F}_{uuu} &=& -4 J^{(1)F}_{u} A_u^2-4 J^{(2)F}_{uu} A_u+\left(\frac{8 A_u^3}{3}-
\frac{A_u T}{3}\right) \hbar +\frac{T J^{(1)F}_{u}}{6}+j^{(3)F}_{uuu}  \, , \nn
J^{(4)F}_{uuuu} &=&
+\hbar  \left(4 A_u^4-\frac{7 T A_u^2}{5}-\frac{2}{5} \left(\nabla _u^2A_u\right) A_u+
\frac{7 T^2}{240}+\frac{3}{5} \left(\nabla _uA_u\right)^2\right)
\nn
&&-8 J^{(1)F}_{u} A_u^3-12 J^{(2)F}_{uu} A_u^2+\left(\frac{1}{5} \nabla _u^2J^{(1)F}_{u}+
\frac{7 T J^{(1)F}_{u}}{5}-6 J^{(3)F}_{uuu}\right) A_u-\frac{3}{5} 
\left(\nabla _uA_u\right) \left(\nabla _uJ^{(1)F}_{u}\right) \nn 
&&+\frac{1}{5} \left(\nabla _u^2A_u\right) J^{(1)F}_{u}+\frac{7 T J^{(2)F}_{uu}}{10}+
j^{(4)F}_{uuuu} \, , \nn
J^{(5)F}_{uuuuu} &=& 
\hbar  \left( \frac{32 A_u^5}{5}
-\frac{104 T A_u^3}{21}-\frac{16}{7} \left(\nabla _u^2A_u\right) A_u^2+\frac{27 T^2 A_u}{70}+\frac{24}{7} \left(\nabla _uA_u\right)^2 A_u
\right. \nn && \qquad \left. 
+\frac{1}{35} \left(\nabla _u^2T\right) A_u-\frac{1}{7} \left(\nabla _uA_u\right) \left(\nabla _uT\right)
+\frac{2}{21} T \left(\nabla _u^2A_u\right)\right)
\nn&&
-16 J^{(1)F}_{u} A_u^4-32 J^{(2)}_{uu} A_u^3+\frac{8}{7} \left(\nabla _u^2J^{(1)}_{u}\right) A_u^2+
\frac{52}{7} T J^{(1)F}_{u} A_u^2-24 J^{(3)}_{uuu} A_u^2
\nn &&
-\frac{24}{7} \left(\nabla _uA_u\right) \left(\nabla _uJ^{(1)F}_{u}\right)
A_u+\frac{12}{35} \left(\nabla _u^2J^{(2)}_{uu}\right) A_u+\frac{16}{7} 
\left(\nabla _u^2A_u\right) J^{(1)F}_{u} A_u
\nn &&
+\frac{52}{7} T J^{(2)F}_{uu} A_u-8 J^{(4)}_{uuuu} A_u 
+\frac{1}{14} \left(\nabla _uT\right) \left(\nabla _uJ^{(1)F}_{u}\right)-\frac{12}{7} \left(\nabla _uA_u\right) \left(\nabla _uJ^{(2)F}_{uu}\right)
\nn &&
-\frac{1}{21} T \left(\nabla _u^2J^{(1)F}_{u}\right)-\frac{27 T^2 J^{(1)F}_{u}}{140}-\frac{12}{7} \left(\nabla _uA_u\right)^2 J^{(1)F}_{u}
\nn &&
-\frac{1}{70} \left(\nabla _u^2T\right) J^{(1)F}_{u}+\frac{8}{7} \left(\nabla _u^2A_u\right) J^{(2)F}_{uu}+\frac{13 T J^{(3)F}_{uuu}}{7}+j^{(5)F}_{uuuuu}~,
\eea
where $A_u$ is a component of the background gauge potential that characterises the motion of fermionic matter in the 
Kerr black hole, which is proportional to the angular momentum quantum number $a$. 
We note that the presence of $\hbar$  in the above formulae is necessary for dimensional reasons, 
and that the propagator of the fermionic field
is given by $\langle \Psi^\dagger (z) \, \Psi (w) \rangle = \hbar (z-w)^{-1}$.

In the limit of the Schwarzschild black hole of interest to us here, 
the gauge field $a \to 0$ and so $A_u \to 0$. Thus the fermionic current expressions become
\bea\label{schfer}
J^{(1)F}_{u} &=& j^{(1)F}_{u} \, , \nn
J^{(2)F}_{uu} &=& \left(-\frac{T}{12}\right) \hbar
+ j^{(2)F}_{uu} \, , \nn
J^{(3)F}_{uuu} &=& \frac{T J^{(1)F}_{u}}{6}+j^{(3)F}_{uuu}  \, , \nn
J^{(4)F}_{uuuu} &=&
+\hbar  \, \frac{7 T^2}{240}  +
\frac{7 T J^{(2)F}_{uu}}{10}+
j^{(4)F}_{uuuu} \, ,  \nn
J^{(5)F}_{uuuuu} &=& \frac{1}{14} \left(\nabla _uT\right) \left(\nabla _uJ^{(1)F}_{u}\right) - \frac{1}{21} T \left(\nabla _u^2J^{(1)F}_{u}\right)-\frac{27 T^2 J^{(1)F}_{u}}{140} -\frac{1}{70} \left(\nabla _u^2T\right) J^{(1)F}_{u} +\frac{13 T J^{(3)F}_{uuu}}{7}+j^{(5)F}_{uuuuu}~.
\eea
We now remark that, as shown in \cite{bonora}, the covariantised versions of the currents  (\ref{wbosonic}, \ref{wfermionic})
with spins higher than two are free of, or at most have trivial,  conformal or diffeomorphism anomalies. 
This is consistent with the fact that the higher moments of the Hawking radiation are expected to describe a
gravitational anomaly-free theory, since only the spin-two current (stress tensor) of the theory
has diffeomorphism or conformal anomalies, and it is the requirement of their
cancellation that requires the appearance of the Hawking radiation 
spectrum~\cite{wilczek}. If these currents had conformal anomalies, 
then they would correspond to new (non-gauge) quantum numbers for black holes, 
which would violate the no-hair theorem. 

The covariant higher-spin-$s$ currents $J^{(s) B,F}_{\mu_1 \mu_2 \dots \mu_n} $ are sourced by 
appropriate background fields ${\mathcal B}^{(s)B,F}_{\mu_1\mu_2\dots \mu_n}$:
\be\label{sources}
J^{(s)B,F}_{\mu_1 \dots \mu_{n}}  = \frac{1}{\sqrt{g}}\, \frac{\delta}{\delta {\mathcal B}^{(s)B,F\, \mu_1 \dots \mu_n} }S \, ,
\ee
where $S$ is the two-dimensional effective action of the Hawking radiation in the near-horizon geometry of the 
spherically-symmetric black hole. The relevant interactions in this effective geometry are then given simply by
\be\label{e2dg}
S_{\rm int} = \int_{\rm near~horizon~2D~space-time} \, d^2 x \sqrt{g}  \, \sum_s \, \sum_{\alpha=B,F} {\mathcal B}^{(s) \alpha\, \mu_1 \dots \mu_n} \, 
J^{(s) \alpha}_{\mu_1 \dots \mu_{n}}~,
\ee
with $x $ denoting two-dimensional space-time coordinates (e.g., in one frame $x=\{ u,v\}$). 
The background fields ${\mathcal B}^{(s) \alpha \, \mu_1 \dots \mu_n}$ may be taken taken to vanish at asymptotic spatial infinity,
far away from the horizon. Due to the quadratic nature of the current, after appropriate partial integrations 
the action (\ref{e2dg}) can be written schematically in the form 
\be\label{covact}
S_{\rm int} = \int_{\rm near~horizon~2D~space-time} \, d^2 x \sqrt{g}  \Big[ {\mathcal V}(x) + \sum_s \, \, \sum_{\alpha=B,F}  \chi^{\dagger \alpha} (x) \, {\mathcal F}^{(s) \alpha} (\partial_\mu) \, \chi^{\alpha} (x) \Big] \, ,
\ee
where $\chi^\alpha $ is a scalar ($\phi(u,v)$ for $\alpha = B$) or fermionic ($\psi (u,v)$ for $\alpha =F$) field, 
and the ${\mathcal F} (\partial_\mu)$ are appropriate functions containing multiple derivatives $\partial_\mu$, $\mu=u,v$ 
with respect to the two-dimensional horizon space-time. The quantity ${\mathcal V}[\varphi (x)]$, 
which is a $\chi$--independent function of the scalar field $\varphi (x)$ and its derivatives, 
plays the r\^ole of a vacuum energy term in the two-dimensional horizon effective field theory. 
It arises from the $\chi$-independent terms of the covariant currents (\ref{Jsjs}), (\ref{schfer}), 
which are generically functions of $\Gamma$ and $T$ (\emph{i.e} of $\partial_u \varphi$, $\partial_u^2 \varphi$)
and their covariant derivatives. 

In the spin-2 case the corresponding spin-2 current (the stress tensor) couples to the graviton
field, $\int d^2 x \sqrt{g} \, T^{\mu\nu} \, g_{\mu\nu}$, which is characterised by diffeomorphism invariance:  
$\delta g_{\mu\nu} = \partial_{(\mu} \xi_{\nu)}$ for an infinitesimal diffeomorphism $\xi_\mu \to x_\mu + \xi_\mu$, 
provided the stress tensor is conserved~\footnote{In the black hole case, as we have discussed above, 
the diffeomorphism invariance is broken by the outgoing flux, but the form of the transformation is included in (\ref{gs}).}.
Generalising this, the higher-spin currents, which are free from conformal and diffeomorphism anomalies~\cite{bonora},  
are conserved exactly, and their conservation is associated with an infinity of Abelian gauge symmetries of the form
\be\label{gs}
{\mathcal B}_{\mu_1 \dots \mu_n}^{(s)} \rightarrow {\mathcal B}_{\mu_1 \dots \mu_n}^{(s)}  + \partial_{(\mu_1 } \, \Xi_{\mu_2 \dots \mu_n)}
\, , \ee
where the $(\dots )$ among indices indicates appropriate symmetrisation. 
The presence of this infinite set of gauge symmetries is consistent with the no-hair theorem, 
as the spatial integrals of the currents correspond to conserved charges. 

The existence of a $W_\infty$ symmetry
of matter in the near-horizon geometry,
larger than the Virasoro algebra, results in the complete integrability of the matter system, 
and is analogous to the cases of matter in the near-horizon geometries of black-hole structures in the
context of string theory~\cite{emn1,zanon}. 
These $W_\infty$ algebras are phase-space-preserving algebras, like the $W_\infty$ algebras 
discussed in the stringy cases earlier. To see this, one may rewrite the (traceless) energy momentum tensor of the 
two-dimensional effective theory using a point-splitting method~\cite{bonora}, as follows (we
consider scalar fields $\phi$ for concreteness): 
\bea\label{tmn}
T_{\mu\nu} &=& {\rm lim}_{y \to 0}\, \partial_\mu \phi (x - y) \partial_\nu \phi (x + y) - g_{\mu\nu}\,  \Big({\rm stress-tensor \, trace} \Big)  \nonumber \\
&=& \sum_{i=0} \, \sum_{j=0} \frac{(-1)^i}{i!\, j!} : y^{\mu_1}\dots y^{\mu_i} \, y^{\nu_1} \dots y^{\nu_j} \, \partial_\mu\, \partial_{\mu_1} \dots \partial_{\mu_i} \phi (x) \, 
\partial_\nu\, \partial_{\nu_1} \dots \partial_{\nu_j} \phi (x) : \, .
\eea
This expression can be covariantised by replacing the partial derivatives by covariant derivatives, 
making the right-hand side of (\ref{tmn}) a complicated expression in terms of products of the  higher-spin currents 
discussed above with  the $y$-dependent factors (\ref{tmn}), that correspond to higher-level background tensors 
$\mathcal B^{(s)}_{\mu_1 \dots \mu_{i_n}}$ that source the higher-spin currents. For our purposes, 
the most important feature of (\ref{tmn}) is the fact that the right-hand side depends not only on the coordinate 
$x^\mu$ but also on the coordinate $y^\mu=dx^\mu$ of the cotangent bundle, 
and thus on a symplectic phase-space manifold, showing that the corresponding $W_{1+\infty}$ algebra 
generated by the higher-spin currents of the Hawking radiation spectrum is indeed a phase-space algebra. 

\section{Field Theory of Hawking Radiation and Horizon-Area-Preserving Classical Symmetries \label{sec:preserve}}

In this Section we demonstrate that one may associate $W$ symmetry with classical
horizon-area-preserving diffeomorphisms,
following the discussion of~\cite{emn2015} for the SU($\infty$)-coloured black-hole case.
To this end, we first consider the completely integrable field theory system of bosonic (scalar) field currents (\ref{wbosonic})
in a flat space-time, and then generalise it to the curved space-time case (\ref{Jsjs}). 

We consider a holomorphic, {i.e.}, one-dimensional, Euclideanised scalar field theory given by the appropriate flat space-time 
limit of (\ref{e2dg}), which involves only the bosonic higher-spin currents $j^{(s)}_{\mu_1\dots \mu_n}$
that are quadratic in the fields $\phi(z)$:
\be\label{holft}
S_{\rm int} = \int_{\rm near-horizon~2D~space-time} \, dz  \, \sum_{s}  {\mathcal B}^{B\, (s) \, z \dots z} \, 
j^{B\, (s)}_{z \dots z}~,
\ee
and we take ${\mathcal B}^{B\, (s)\,  z \dots z}$ to be asymptotically constant. Note that these background fields are \emph{not} 
functions of the $\phi(z)$ fields, but may be functions of the holomorphic coordinate $z$.
Introducing the Fourier transforms of the fields 
\be\label{fourier}
\phi(z) = \int dp \, e^{i p\, z} \, {\tilde \phi}(p) \, ,
\ee
and defining 
\be\label{matrixfields}
{\mathcal U}_{pz} \equiv {\tilde \phi}(p)\, e^{ip\, z} \ne {\mathcal U}_{zp} \, ,
\ee
we observe that the quantity 
\be\label{upzsum}
\int_{-\infty}^\infty dz\, \int_{-\infty}^\infty dp \, U_{pz} = 2\pi \, {\tilde \phi}(0)
\ee
is proportional to ${\mathcal U}_{0z}$. Thus, if we view the indices $p,z$ as spanning a set of \emph{discrete values}: 
$ 0, 1, \dots N-1$ with $N \to \infty$ (the set becoming continuous in the limit only), we observe that, on account of the 
constraint (\ref{upzsum}) the (complex) field variables ${\mathcal U}_{pz}$ have only $N^2 - 1$ independent 
degrees of freedom, as $N \to \infty$. If we label these degrees of freedom as 
\be\label{variables}
{\mathcal U}_{pz} \rightarrow f_\alpha, \, \quad  \alpha=1, \dots N^2-1~,
\ee 
in which case the interacting action (\ref{holft}) becomes 
\be\label{holft2}
S_{\rm near~horizon~scalars} = \sum_{\alpha,\beta}  \, f_\alpha \, {\mathcal C}^{\alpha\beta} \, f_\beta~, \qquad \alpha, \beta =1, \dots N^2-1~,
\ee
and the (matrix) coefficients ${\mathcal C}^{\alpha\beta}$ contain terms $z^m \, p^n$, with $m, n$ positive integers, 
where (in operator form) $p=-\partial_z$. 
 
There is a local (gauge) symmetry characterising the action (\ref{holft2}) since, in the spirit of \cite{wcontr}, 
one may redefine the field variables by unitary matrices 
 ${\mathcal V}_\alpha^\beta$:
 \be\label{unitary}
 f_\alpha \rightarrow {\mathcal V}_{\alpha}^{\,\,\,\,\beta}\, f_\beta \, ,
 \ee
and integrate over ${\mathcal V}$ in a path integral. The original action (\ref{holft2}) may then be viewed as a 
``gauge fixed'' version of the theory, where ${\mathcal V}$ is fixed in an appropriate form. 
 
The ``time'' in this Euclideanised near-horizon geometry is not the target time included in the light-cone variable $z$,
but it is the Liouville RG time (\ref{time}), which characterises the evolution of the scalar modes $\phi(z)$. 
From a stringy black hole viewpoint, the latter correspond to propagating modes, and as such they correspond to 
non-marginal deformations in a $\sigma$-model that describes string propagation in the neighbourhood of the 
black hole~\cite{Chlyk,emn1,emn2}. This Liouville time $t$ leads to an one-dimensional adjoint Higgs model for the 
Hawking-radiation scalar matter, as a consequence of the Lagrangian (\ref{cov}), namely 
 \be\label{adjhiggs}
 S_H =\int d t \, \tr\left[ \frac{1}{2}
\big( \del_0 \hat{M} (t) 
- [\hat{\ba}_0 ,\hat{M}](t)\big)
\big( \del^0 \hat{M} (t) 
-[\hat{\ba}^0 ,\hat{M}](t)\big)-v(\hat M)\right] \, ,
\ee
where the index $0$ denotes the time variable, the trace is over group indices, and the matrix-valued field 
$M(t)=f_\alpha \, T^\alpha$, with $\alpha=1, \dots N^2 -1$, and
the $(N^2-1) \times (N^2-1)$ matrices $T^\alpha$ form an adjoint representation of the SU($N \to \infty$) gauge group. 
The properties of such a theory connected with the $W_\infty$ gauge symmetries are discussed in the Appendix, 
following the analysis of \cite{wcontr}. 
The fact that  we are using an adjoint  matrix representation of the SU($\infty$) algebra is important, 
because - rigorously speaking~\cite{uehara} -  it is the large-$N$ limit of the  $(N^2-1) \times (N^2-1)$ matrix 
generators in the adjoint representation of the SU(N) gauge group that become those of the Poisson algebra in the classical limit. 
In contrast, the $N \times N$ matrices in the fundamental representation of SU(N) diverge and are not well defined in the 
$N \to \infty$ limit. The gauge potential $\hat{\ba}_0$ is identified appropriately with the square root of the central charge 
deficit of the world-sheet theory $Q(t)$, according to the discussion leading to (\ref{cov}). The potential in our case 
contains at most quadratic terms in the adjoint Higgs field $\hat{M}$, given the form of the action (\ref{holft2}). 
 
The area-preserving nature of the associated classical $w_\infty$ symmetries is seen by viewing the action 
(\ref{adjhiggs}) as a ``gauge'' theory over 
an extended $(2 + 1)$-dimensional space-time, where the ``internal'' space is viewed as a stereographic projection 
of a spherical surface ($z, \overline z$)
representing the black hole horizon of a macroscopic (semi-classical) black hole. In such a large-area black hole, 
the horizon can be approximated by an almost flat surface, and hence the approach~\cite{wcontr} of constructing 
$W_\infty$ gauge theories and their classical limit outlined in the Appendix applies.
The corresponding adjoint Higgs action is given by (\ref{higgs}) with $d=1$, and the local limit (\ref{rescaling})
yields an invariance of the corresponding limiting action (\ref{localYM}) under the horizon-area-preserving  diffeomorphisms (\ref{prtrns}). 

We turn next to the fermionic currents in the near-horizon geometry, which (like the bosons) can also be represented 
as a one-dimensional quantum-mechanical system, by labelling the holomorphic fermions as 
\be\label{fermions}
\Psi (z,t) = \psi_{z}(t) \, ,
\ee
with $t$ denoting the Liouville RG time. 
As in the bosonic case, we view the (continuous) suffix $z$ as an internal fermion index, 
a limiting case of a discrete index: $\tilde z=1, \dots N$, $N \to \infty$.
In this case, any integration over $z$ becomes a sum over internal fermion indices: $\int dz \, \sum_{\tilde z}$. 
Likewise, any determinant of the two-dimensional black-hole metric and other functions of the original holomorphic 
variables $z$ become functions of the (discretised) $\tilde z$. 

Using this representation and including the RG ``time'' dependence of the fermion fields, 
the fermionic conformal field theory near the black-hole horizon, with interaction terms (\ref{e2dg}), 
becomes a quantum-mechanical theory (field theory in $d=1$ space-time dimension)
of the fermions $\psi_z (t)$  (\ref{fermions}) of the generic form:
\be\label{fermilagrange}
{\mathcal S}_\psi = \int dt \sum_{\tilde z} \psi_{\tilde z}^\dagger (t) \, \Big( i\, \frac{d}{dt}  +  h^{(1)}[\partial_{\tilde z}]\Big) \, \psi_{\tilde z} \, ,
\ee
where the structure $h^{(1)}[\partial_z]$ represents the complicated operators in the interaction terms (\ref{e2dg}) 
after appropriate partial integration. (We assume that the background source fields ${\mathcal B}_{\mu_1 \dots \mu_n}$
are static, and the upper index (1) in  $h^{(1)}$ indicates that it pertains to fermion bilinear terms only.) 
The gauge nature of the theory can be seen by 
observing that such constructions hide an infinite-dimensional gauge theory~\cite{wadia,wcontr} (a ``gauged'' $w_\infty$ algebra). 
of the fermion transformations
\be\label{gtrnsf}
\psi_{\tilde z} \to U_{\tilde z}^{\,\, {\tilde z^\prime}}  \, \psi_{\tilde z^\prime} \, ,
\ee
where $U_{\tilde z}^{\,\, {\tilde z^\prime}}$
is an $N \times N$ {unitary} matrix, in the fundamental representation of the $SU(N \to \infty)$ group,
as opposed to the adjoint representation in the bosonic case.

The proper gauging procedure for the associated $W_\infty$ algebra pertaining to the fermion case
is also presented in the Appendix, where the fermion
action (\ref{fermilagrange}) is represented as a (1+2)-dimensional action over the extended (horizon) surface $z, \overline z$.  
However, because the fermions are in the fundamental representation, the appropriate Poisson large-$N$ limit 
cannot be defined~\cite{uehara}. Indeed,
according to the discussion in the Appendix, upon taking the local limit $\ell \to 0$ in (\ref{rescaling}), 
by means of which one defines a classical contraction $w_\infty$ of the quantum $W_\infty$ algebra, 
one obtains a trivial fermion action ${\mathcal S}_\psi \to 0$, as $\ell \to 0$. 
Thus the area-preserving nature of the classical $w_\infty$ symmetries that characterise the black hole horizon is realised 
non-trivially via the bosonic Hawking radiation fields.

The above results have been obtained in the flat space-time limit. Nevertheless, covariantising the flat-space current theory
and going to a curved space-time metric using (\ref{Jsjs}) does not change qualitatively the above features of the flat 
space-time theory, as can readily be seen from the form of the covariant effective action (\ref{covact}). 
In a phase-space representation, the 
function ${\mathcal F}^{(\alpha)}(\partial_\mu)$  will still play the r\^ole of a hamiltonian operator as in (\ref{holft2}) and
(\ref{fermilagrange}), and the only remnant of the curved metric would be the vacuum energy term 
$\int d^2 x \sqrt{-g} {\mathcal V}[\varphi]$, which is invariant under the gauge $W$-symmetries, 
being field $\phi,\psi$--independent~\footnote{The metric field $\varphi (u,v)$, corresponding to the conformal 
factor of the near-horizon metric in the particular conformal-frame representation (\ref{nhg}) 
of the two-dimensional horizon geometry, does not transform under the gauge symmetries in question.}. 

The area-preserving {classical} $w_\infty$ symmetries are consistent with the view of the horizon area of a 
classical black hole as a conserved Noether charge~\cite{wald}. On the other hand, at a quantum level, 
the $W_\infty$ quantum symmetries, although they are phase-space area-preserving symmetries for matter in the near-horizon 
geometry that maintain the complete integrability of the matter system (and hence preserve quantum coherence~\cite{emn1})
do not preserve the horizon area. This feature is in agreement with the shrinking of the latter quantity
with increasing time, due to the Hawking evaporation process. 

\section{Conclusions and Outlook \label{sec:concl} }

We conclude by re-iterating the main points underlying the microscopic mechanism for maintaining 
quantum coherence and retaining information in an evaporating spherically-symmetric stringy black hole. 
The corresponding effective theory is a two-dimensional string theory, with a
singularity structure whose dynamics is described by an integrable physical system characterised by an infinity of 
mutually-commuting $W_{1+\infty}$ conserved charges, carried by non-propagating delocalised discrete gauge states, 
corresponding to an infinity of higher-spin states. These discrete states have zero norm and discrete momenta, 
which, however, take on the same values as those corresponding to the infinity of physical (positive norm) 
propagating string states of the effective two-dimensional string.
The two-dimensional substructure is essential to this argument, and can always be embedded in four dimensions 
by considering near-horizon geometries of the form SU(1,1)/U(1) $\times \, H^{(2)}$, 
where $H^{(2)}$ a two-dimensional compact or non-compact manifold. 

In this picture, the infall of matter into the black hole horizon and the Hawking radiation process are viewed as `particle interactions' in the following sense.
Consider first the case of matter falling into this black hole,
specifically massless matter (represented as a `tachyon' propagating mode in the above effective-two-dimensional 
string theory context), which starts from spatial infinity. Initially, world-sheet conformal invariance of this tachyon 
background is guaranteed without mixing with the higher-spin states. 
However, upon reaching the horizon, discrete delocalised string modes of higher spin are excited,
in order to dress the tachyon background and make it conformal on the world-sheet. In this sense, 
given that the $W_\infty$-charges are conserved, the back reaction of the black hole
leads to an excited state, so that the whole process can be represented as
(stringy black hole) + (massless~matter) $\Rightarrow$ (stringy black hole)$^\star$,
where the star denotes an excited state and a rearrangement of the $W_\infty$-charges. 
The black hole is viewed as a string state in the background of discrete gauge states and other topological states in this picture, 
as per our description above and in the previous literature. 

The Hawking evaporation process can be thought of similarly as successive steps of the time-reversed process. 
This is reminiscent of the arguments of \cite{visser} for viewing black holes as `particles' and their Hawking radiation 
as consisting of successive  `two-body' decays, in accordance with the sparsity of the Hawking radiation at infinity. 
However, our picture is very different in essence, as the black holes are stringy states characterised by an infinity of charges, 
thus integrable systems. The emitted massless matter reaches spatial infinity ``decoupled''  (in the sense
that its world-sheet $\beta$-function vanishes) from the topological states,  
but the latter (due to conservation of the $W_\infty$ charges) are omnipresent as a non-thermal discrete environment, 
carrying information.  In string theory, the thermal Hawking radiation spectrum is only `part of' the whole picture, 
associated with propagating modes, whereas the discrete states provide a specific `Ariadne's thread'
for external measurements capable~\cite{wmeasure} of
reconstructing the information `mislaid' within the black-hole labyrinth.

Comparing finally with the supertranslation approach~\cite{strominger, strominger2,HPS}, 
we recall that the latter also lead to an infinity of conserved charges on the two-dimensional horizon, 
which correspond to currents excited during the interaction of infalling matter. 
This can be seen straightforwardly in the representation of the black hole horizon as a recoiling D-brane~\cite{emn2015}. 
However, the supertranslation charges are not responsible for balancing the information books, 
for the reasons stated above. This role is played by the $W_\infty$-charge-carrying topological discrete states,
as can be seen in the simple example of the two-dimensional black hole~\cite{witt}, 
where these states are responsible for retaining information even in the absence of an horizon,
which is only a spatial point in two dimensions.

\section*{Acknowledgements}

The work of J.E. and N.E.M. was supported in part by the London
Centre for Terauniverse Studies (LCTS), using funding from the European Research Council via the
Advanced Investigator Grant 267352 and by STFC (UK) under the research grant ST/L000326/1, while that of D.V.N. 
is supported in part by the DOE Research Grant DE-FG02-13ER42020.

\section*{APPENDIX: $W_\infty$ Symmetries in $d$ Space-Time Dimensions as $d+2$-Dimensional Gauge Symmetries}

We discuss in this Appendix the connection between the Hawking radiation fields on the horizon of the 
spherically-symmetric black hole with classical SU($\infty$) area-preserving gauge symmetries. 

We consider the construction~\cite{wcontr} of quantum $W_\infty$ (and classical $w_\infty$) gauge theories in 
$(d+2)$ dimensions, 
where $d$ is the dimension of the space-time where the algebras live: $d=2$ in the black-hole case of interest to us. 
The $W_\infty$ quantum algebra may be defined as the
algebra of commutators of Hermitian operators $\xi (a, a^\dagger)$, where $a, a^\dagger$ are harmonic-oscillator 
annihilation and creation operators. 
The operators $\xi(a, a^\dagger)$ may be parameterised using coherent states on a Euclidean space~\cite{wcontr}, 
spanned by complex coordinates 
$z$, $\overline z$, which can be identified with the coordinates of a stereographic projection of the 
horizon sphere $S^2$~\cite{emn2015}:
$$ :\xi (\hat a, {\hat a}^\dagger): = \int d^2z \, e^{-|z|^2} \, |z> \, \xi (z, \overline z) \, <z|~,  $$
where $|z> = e^{{\hat a}^\dagger \, z} \, |0>, \, <z| = <0| \, e^{\hat a \, \overline z}, \, <z^\prime | z > = e^{{\overline z}^\prime\, z}, \, \hat a\, |z> = z \, |z>,  \, <z| \, {\hat a}^\dagger = < z|\, \overline z$, we use
the normalisation condition $\int d^2 z\,e^{-|z|^2} |z>\,<\overline z| = 1$ with $d^2 z \equiv  \frac{1}{\pi} {\rm Re}\, z \, {\rm Im} \, z$, 
and $:\xi(\hat a, {\hat a}^\dagger):$  is a(n anti-)normal-ordered operator, 
in which the annihilation operators are always placed to the left of the creation operators. 

One may regard~\cite{wcontr} the coordinates $z, \overline z$ as a group-theoretical (`colour') space, and
introduce a gauge potential $A_\mu (x, \hat a, {\hat a}^\dagger) $, 
where $\mu = 1 , \dots d$ is a $d$-dimensonal space time $\{ x \}$ index:
\be\label{gp}
\hat A_\mu (x) \equiv A_\mu (x, \hat a, {\hat a}^\dagger) = \int d^2 z \, e^{|z|^2} \, |z> \, A_\mu (x, z, \overline z) \, < z| \, .
\ee
One may then introduce an infinite-dimensional set of infinitesimal $W_\infty$ gauge transformations:
\be\label{gtr}
\delta \hat A_\mu (x) = \partial_\mu \hat \xi (x) + i \Big[\hat \xi (x), \, \hat A_\mu(x) \Big]~, \quad
\delta A_\mu (x, z, \overline z) = \partial_\mu \xi (x, z, \overline z) - \{\{ \xi, \, A_\mu \}\}_{\rm Moyal}(x, z, \overline z)~,
\ee
where the symbol $\{\{ ., . \}\}_{\rm Moyal}$ denotes a Moyal bracket, defined by: 
\be\label{moyal}
\{\{ \xi_1, \, \xi_2 \}\}_{\rm Moyal}(z, \overline z)  \equiv i \sum_{n=1}^{\infty} \frac{(-1)^n}{n\!} \Big( \partial_z^n \xi_1 (z, \overline z) \, 
\partial_{\overline z}^n \, \xi_2 (z, \overline z) - \partial_{\overline z}^n \xi_1 (z, \overline z) \, 
\partial_{z}^n \, \xi_2 (z, \overline z)\Big) \, .
\ee
The generators of $W_\infty$, $\rho [\xi ]$, are linear functionals of $\xi (z, \overline z)$ in this construction, satisfying
\be 
\Big[\, \rho[\xi_1], \, \rho[\xi_2]\Big] = i \rho[\{\{ \xi_1, \, \xi_2 \}\}_{\rm Moyal}]
\ee
at the quantum level~\cite{wcontr}.
The classical area-preserving $w_\infty$ Lie algebra, as obtained from $W_\infty$ by the appropriate contraction 
discussed in \cite{wcontr},  is then
\be 
\Big[\, \rho[\xi_1], \, \rho[\xi_2]\Big] = i \rho[\{ \xi_1, \, \xi_2 \}_{\rm Poisson}] \, ,
\ee
where $\{., \, .\}_{\rm Poisson}$ denotes the (classical) Poisson bracket. 

We observe that, in this representation, the $W_\infty$ gauge fields $A_\mu(x, z, \overline z)$ 
are defined with coordinates in a $d+2$-dimensional space-time
$\{x,\, z, \,\overline z\}$ with a two-dimensional `internal' `colour' space spanned by the $\{z, \, \overline z\}$ coordinates. 
We consider the following Yang-Mills-type action $\mathcal S$, which is invariant  under the $W_\infty$ 
gauge transformations (\ref{gtr}):
\begin{equation}\label{YMahat}
\mathcal S = -\frac{1}{4g^2} \, \int d^d x \frac{1}{4} {\rm Tr} \Big(\hat {\mathcal F}_{\mu\nu}\, \hat {\mathcal F}^{\mu\nu} \Big)~:~
\hat {\mathcal F}_{\mu\nu} = 
\partial_\mu \hat A_\nu (x) - \partial_\nu \hat A_\mu (x)  -i  \Big[\hat A_\mu, \, \hat A_\nu \Big]~, 
\end{equation}
where $g$ is a coupling constant. This can be rewritten  using the coherent-state representation as~\cite{wcontr}:
\bea\label{YM}
\mathcal S &=& -\frac{1}{4g^2}\,  \int d^d x d^2 z \, \sum_{n=0}^{\infty}\, \frac{(-1)^n}{n\!} \, \partial^n_z {\mathcal F}_{\mu\nu}(x, z, \overline z) \, \partial_{\overline z}^n {\mathcal F}^{\mu\nu}(x, z, \overline z)~: \nonumber \\
{\mathcal F}_{\mu\nu} &=& \partial_\mu A_\nu (x, z, \overline z) - \partial_\nu A_\mu (x, z, \overline z)  + \{\{ A_\mu, \, A_\nu \}\}_{\rm Moyal}(x, z, \overline z)~.
\eea
We note that the action is non-local in terms of the $z,\overline z$ variables. 
Indeed, as stressed in \cite{wcontr},
it is this non-local nature of the action that differentiates the quantum $W_\infty$ from the classical $w_\infty$ symmetry,
as concerns the association with the SU($\infty$) gauge theory.
It is the $W_\infty$ that can be viewed as the $N \to \infty$ limit of SU(N), not the classical $w_\infty$.

Next, we consider a scalar field in an adjoint representation, which 
we call a Higgs field:
\be
\hat M (x)\equiv M(x,\hat a,\hat a ^\dag) =\int\zintm\ve z \ke M
 (x,z,\zbar)\br z\ve \, .
\label{higgsfield}
\ee
The Yang-Mills-Higgs action is then given by:  
\bea 
&&
S_H =\int d^d x \tr\left[ {1\over 2}
\big( \del_\m \hat{M} (x) 
- [\hat{\ba}_\m ,\hat{M}](x)\big)
\big( \del^\m \hat{M} (x) 
-[\hat{\ba}^\m ,\hat{M}](x)\big)-v(\hat M)\right] \nn
&& \hspace{-0.5cm} =
\int d^d x \left[ \int d^2 z 
{1\over 2}{\sum_{m=0} ^{\infty}}{{(-)^m}\over{m!}}
{\partial_{z} ^{m}}\big( \del_\m M (x,z, \zbar) 
-\{\!\!\{\ba_\m ,\M\}\!\!\}(x,z,\zbar)\big)\times 
\times {\partial_{\bar{z}} ^{m}} 
\big( \del^\m M (x,z, \zbar)  
-\{\!\!\{\ba^\m ,\M\}\!\!\}(x,z,\zbar)\big) \right]
-\tr  v(\hat M),\nn
&& v(\hat M) = \sum_n g_n {\hat M}^n, \ \ \ \ \ \ \ \ \ 
dim(g_n) = d\left({n \over{2}} - 1\right) - n. 
\label{higgs}
\eea
We require that the fields and their 
$z,\zbar$ derivatives should fall to zero 
at $z=\inf$. It is easy to check that this action is invariant under the
\Winf gauge transformation (\ref{gtr}) and 
\be
\d M (x,z,\zbar)= \{\!\!\{\x,M \}\!\!\}(x,z,\zbar)~,
\quad
\d \hat M (x)=-i[\hat\x (x) ,\hat M (x)] .
\label{higgsgauge}
\ee
Notice that here again the interactions are 
non-local in the internal ($z, \overline z$) space.

As a last example of a \Winf gauge theory, let us introduce a fermion field in a fundamental representation 
of \Winf, namely a field that transforms as a bra or ket vector in the Hilbert space of a harmonic oscillator:
\bea
\ve \y (x)\ke =\int \ve z \ke \zintm\br z\ve \y (x)\ke\equiv\int
\ve z \ke \  \zintm \y (x, \zbar ) \, .
\eea
We can write the action as 
\bea\label{fermions2}
S_F & = &\int d^n x \br \y (x) \ve \
\g^\m\big(i\del_\m - \hat{\ba}_\m (x)\big)\ve \y (x) \ke
\nn
& = & \int\int d^n x \zintm \bar{\y} 
(x, z)\g^\m\big(i\del_\m -\ba_\m (x, z,\zbar )\big)\y (x, \zbar ) \, ,
\eea
which is invariant under the $W_\infty$ gauge transformation and
\bea
\d \ve \y(x) \ke =  - i \hat\x (x) \ve \y(x) \ke \, ,
\ \ \ \ \ \ \ \ \ \ \ \ \ \
\d\y (x, \zbar ) =-i\ddag\x (\del_{\zbar} ,\zbar )\ddag\y (x, \zbar ) \, ,
\eea
where $\ddag\ \dots \ddag$ indicates that the derivatives 
are placed on the left of $\zbar$.

To consider the classical limit of the gauge algebra, as appropriate for the horizon-area-preserving symmetry 
of the classical black hole, where the horizon area is viewed as a Noether charge~\cite{wald},
one defines the variables $\sigma_i$, $i=x,y$:
 \be\label{classical}
 z=\frac{1}{\sqrt{2}\, \ell} \Big( \sigma_x + i \sigma_y\Big), \quad \overline z=\frac{1}{\sqrt{2}\, \ell} \Big( \sigma_x - i \sigma_y\Big) \, ,
 \ee  
 and take the limit of the length $\ell \to 0$.
 As is well known~\cite{wcontr}, this limiting procedure will yield the contracted classical area-preserving 
 (in the internal ($z,\overline z$) space) $w_\infty$ symmetry from the quantum $W_\infty$ symmetry. 
 For fields in the adjoint representation of the SU($N$) group, such as the gauge field $\ba_\mu$ and the adjoint 
 Higgs $M$ field, this procedure is straightforward. To see this, we first pass from the $z,\overline z$ variables
 to the $\sigma_i$, $i=x,y$ variables in the effective actions above,
and then consider the limit $\ell \to 0$, while performing the simultaneous rescaling of the various fields:
\bea\label{rescaling}
\ba_\mu (x, z, \overline z)&=& \ell^{-2} \, \ba_\mu (x, \sigma_x, \sigma_y), \quad M(x, z, \overline z) = \sqrt{2\,\pi} \ell^{-2} \, M(x, \sigma_x, \sigma_y)~, \nonumber \\
g^2 &=& {\tilde g}^2 \, \ell^{-6}~, \qquad g_n = {\tilde g} \, \sqrt{2\, \pi} \, \ell^{2-n}~.
\eea
The actions (\ref{YMahat}) and (\ref{higgs}) then become: 
\bea\label{localYM}
&& S_{YM} = -{1\over{4\tilde{g}^2}}\int d^d x d^2 \vec{\s} 
F_{\m\n}(x,\vec\s)F^{\m\n}(x,\vec\s) \, : \nonumber \\
&&  F_{\m\n}(x,\vec\s)=\del_\m\ba_\n(x,\vec\s)
-\del_\n\ba_\m(x,\vec\s)  
+\e^{ij}\del_i\ba_\m(x,\vec\s) \del_j\ba_\n(x,\vec\s) \, , \nonumber \\
&& S_H =
\int d^d x d^2 \vec\s \left[ {1\over 2}
\big(\del_\m M(x,\vec\s) 
- \e^{ij}\del_i\ba_\m(x,\vec\s)\del_j M(x,\vec\s) \big)
\right.\times \nonumber  \\
&&
\left. \times \big(\del^\m M(x,\vec\s) 
- \e^{ij}\del_i\ba^\m(x,\vec\s) \del_j M(x,\vec\s) \big)
- \tilde{v} (M)\right] \, : \nonumber \\ 
&&
\tilde{v} (M) = \sum_n \tilde{g}_n M^n(x,\vec\s) \, .
\eea
Here again we require 
that the fields vanish at $\vec\s = \inf$.

Setting $\x(x,z,\zbar)=l^{-2}\x(x,\vec\s)$, we find 
the \winf gauge transformations: 
\bea
\d \ba^\m(x,\vec\s) &=& \del ^\m\x(x,\vec\s)  
-\e^{ij}\del_i\x(x,\vec\s) \del_j\ba^\m(x,\vec\s) \, , \nonumber \\
\d M(x,\vec\s) &=& \e^{ij}\del_i\x(x,\vec\s) \del_j M(x,\vec\s) \, .
\label{prtrns}
\eea
One can check that
the actions (\ref{localYM}) are indeed invariant under the (classical) 
$w_\infty$  gauge transformation (\ref{prtrns}). The reader should notice 
that the second equation
of (\ref{prtrns}) can be written as 
\bea\label{areaclassical}
\d M(x,\vec\s) = M(x,\vec\s +\d\vec\s (x,\vec\s ))
-M(x,\vec\s ),\ \ \ \ \ \ 
\d\s^i (x,\vec\s ) =-\e^{ij}\del_j\x(x,\vec\s) \, ,
\eea
which is a local
area-preserving coordinate transformation in the internal two-dimensional space. 
 
It is important to stress that he damping factor $e^{-|z|^2}$ 
cancels out in Lagrangians for the fields in the 
adjoint representation such as $\ba_\m$ and $M$,   
due to the property of the trace in the coherent-state 
representation. This allows the limit $\ell \to 0$ in the change of variables (\ref{rescaling}) to be well-behaved, 
yielding non-trivial actions (\ref{localYM}) in that limit.

This is a feature of fields in the adjoint representation of SU($\infty$). The same cannot be said for the 
fermion fields in (\ref{fermions2}), which belong to the fundamental representation of SU($\infty$). 
For the latter action there are no damping $e^{-|z|^2}$ factors in the $d+2$ extended space-time, 
which implies that, formally, the fermion action vanishes in the classical $\ell \to 0$ limit (\ref{rescaling}), $S_F \to 0$.

\end{document}